%% file: manuscript.tex
\documentclass[journal]{IEEEtran}

\usepackage{ucs}
\usepackage[utf8x]{inputenc}
\usepackage[cmex10]{amsmath}
\usepackage{cite, amsfonts, amssymb, amsthm, bm, bbm, graphicx, relsize, multirow, booktabs, color, colortbl,url,soul}
\usepackage[american]{babel}
\usepackage[T1]{fontenc}
\usepackage{algorithmic, algorithm}
\usepackage{tracklang}
\usepackage{glossaries}
\usepackage[tight,TABTOPCAP]{subfigure}

\theoremstyle{definition}

\theoremstyle{remark}

\DeclareMathOperator*{\argmax}{\arg\;\max}

\newcommand{\test}{\mathrel{\underset{H_0}{\overset{\bar{H}_0}{\gtrless}}}} 

\newcommand{\e}{\mathrm{e}}
\renewcommand{\i}{\mathrm{i}}

\graphicspath{{figures/}}

\title{Adaptive detection and localization exploiting \\ the IEEE~802.11ad standard}
\author{Emanuele~Grossi,~\IEEEmembership{Senior Member,~IEEE,} Marco~Lops,~\IEEEmembership{Fellow,~IEEE,}\\Luca~Venturino,~\IEEEmembership{Senior Member,~IEEE} %
\thanks{Emanuele~Grossi and Luca~Venturino are with the DIEI, Universit\`a degli Studi di Cassino e del Lazio Meridionale, Italy 03043; e-mail: e.grossi@unicas.it, l.venturino@unicas.it.} %
\thanks{Marco~Lops is with the DIETI, Universit\`a degli Studi di Napoli ``Federico II,'' Italy 80138; e-mail: lops@unina.it.}
}

\input{acronyms}
\begin{document}
\bstctlcite{BSTcontrol}
\maketitle

\begin{abstract}
In this work, we exploit the sector level sweep of the IEEE 802.11ad communication standard to implement an \emph{opportunistic} radar at mmWaves and derive an adaptive procedure for detecting multiple targets (echoes) and estimating their parameters. The proposed detector/estimator extracts the prospective echoes one-by-one from the received signal, after removing the interference caused by the previously detected (stronger) targets. Examples are provided to assess the system performance, also in comparison with the canonical matched-filter peak-detector and the Cram\'er-Rao bounds. Results indicate that the proposed method is robust against the signal spillover and the near-far problem caused by the imperfect auto-correlation of the probing signal and, for the same probability of false alarm, grants detection and localization performances close to those previously obtained in a simplified single-target scenario.
\end{abstract}


\section{Introduction}
The key enabling technologies of terrestrial wireless communication standards from High Speed Packet Access+ (HSPA+) onwards are based on the availability of a larger and larger transmission bandwidth~\cite{Andrews,Niu2015}, whereby the migration towards carrier frequencies from the S-band (2-4 GHz) upwards was foreseeable long before the establishment of the IEEE 802.11ad standard, operating in the V-band (40-75 GHz)~\cite{802.11adStandard}. The overcrowding of the 2-4 GHz bandwidth (and prospectively of the 4-8 GHz as well), traditionally assigned to radar surveillance and tracking, has pushed the academic and industrial research towards the study of co-existing  architectures wherein radar and communication functions can be both implemented~\cite{Griffiths}: among the many philosophies emerged so far, the one of interest here is the Dual Function Radar Communication (DFRC)~\cite{Bocquet10,Chiriyath16,Hassanien16,Hassanien}, putting forth the idea of a {\em merely functional} coexistence which produces no mutual interference between the sensing and the communication systems. 

In the present contribution, we define {\em opportunistic sensing} the exploitation of communication signals, enabled by a strict coordination between the communication transmitter and a co-located radar receiver chain, to the end of target detection and localization: such a strategy is reminiscent of passive radar systems~\cite{Griffiths-Baker-1,Griffiths-Baker-2}, but is substantially different due to the aforementioned heavy information exchange. Opportunistic sensing has a long history (see, e.g., \cite{Sturm} and references therein) and is still the subject of vibrant research. Not until recently has the use of millimeter waves been proposed to exploit the V-band for communication~\cite{ChannelModelIII,Rappaport13,RappaportBook14,Rappaport_2015}; the key requisite at mmWaves is the establishment of a proper beam alignment between a pair of communicating nodes~\cite{Aryanfar_2014}. For example, the IEEE 802.11ad standard operating at $60$ GHz supports beam steering towards up to 128 distinct sectors to cope with the severe channel attenuation~\cite{802.11adStandard, 802.11ad-com-mag, ABI802ad} and implements a periodic search procedure, composed of a preliminary \gls{sls} phase aimed at acquiring a coarse-grain antenna sector configuration and  by a beam refinement phase, to establish a highly directional communication~\cite{802.11ad-com-mag}. 

The works in~\cite{Gonzalez-conf,Gonzalez-paper} have recently proposed to exploit the single-carrier physical layer frame of IEEE 802.11ad to extract the relevant parameters from a reflecting object in front of the transmitter once a directional link has been established. Instead, the works in~\cite{Grossi-RADAR-2017,Grossi-VTC-2017,Grossi-TSP-2018} have leveraged  the \gls{sls} phase of IEEE 802.11ad to implement short-range surveillance radar, patrolling an entire angular sector; the fundamental limits, in terms of detection probability and range-Doppler localization accuracy, achievable by using different segments of the \gls{cphy} packet transmitted during the \gls{sls} have been established. Monitoring-oriented applications may possibly benefit from these architectures: indeed, there is a growing interest towards performing tasks like collision avoidance, traffic management, intrusion detection, restricted area surveillance, patient monitoring, child and elder home-care without the use of dedicated devices~\cite{Patwari2010, Wilson12, Seifeldin13, Savazzi16, Xu16}. The assumption that most of the above preliminary studies has in common is the presence of a single target in the azimuth sector explored by the transmit beam. 
First efforts to remove this limitation have been undertaken in~\cite{Gonzalez-paper}, but a thorough study of the potentials of 802.11ad radars in a realistic multi-target (i.e., in the presence of an unknown number of possibly moving targets) environment is still lacking. 

In the above context, the contribution of the present study is many-fold. At first, the properties of the \gls{cphy} packet transmitted in the \gls{sls} of the IEEE 802.11ad  communication standard are investigated: indeed, opportunistic sensing cannot exploit one of the major degrees of freedom the radar designer can rely upon, i.e., the transmit waveform, since they are bound to use segments of the communication signal. In particular, the auto-correlation function of the \gls{cphy} packet may present spurious peaks and/or a non-negligible tail that may lead to (weak) target masking and false detections in a canonical \gls{mf-pd} due to the signal spillover. To overcome this limitation, we therefore derive an {\em adaptive} detector-estimator which extracts the prospective echoes one-by-one from the received signal, after removing the interference caused by the previously detected (stronger) targets; here the adaptivity is necessary due to the time-varying nature of the sensed environment, 
Finally, a thorough performance assessment is offered, showing the merits of the proposed approach for short-range patrolling; in particular, multiple possibly moving targets are detected and localized with a precision in the order of few centimeters.

The rest of this paper is organized as follows. Section~\ref{SEC:description} contains the system description, the signal model, and the analysis of the range/Doppler accuracy granted by the available probing signal. Section~\ref{SEC:detector-design} presents the proposed adaptive detector/estimator. Section~\ref{SEC:numerical-results} is devoted to the numerical analysis. Finally, concluding remarks are given in Section~\ref{SEC:conclusions}.

\section{System description}\label{SEC:description}
\begin{figure}[!tp]
	\centering
	\ifCLASSOPTIONtwocolumn
	\includegraphics[width=0.98\columnwidth]{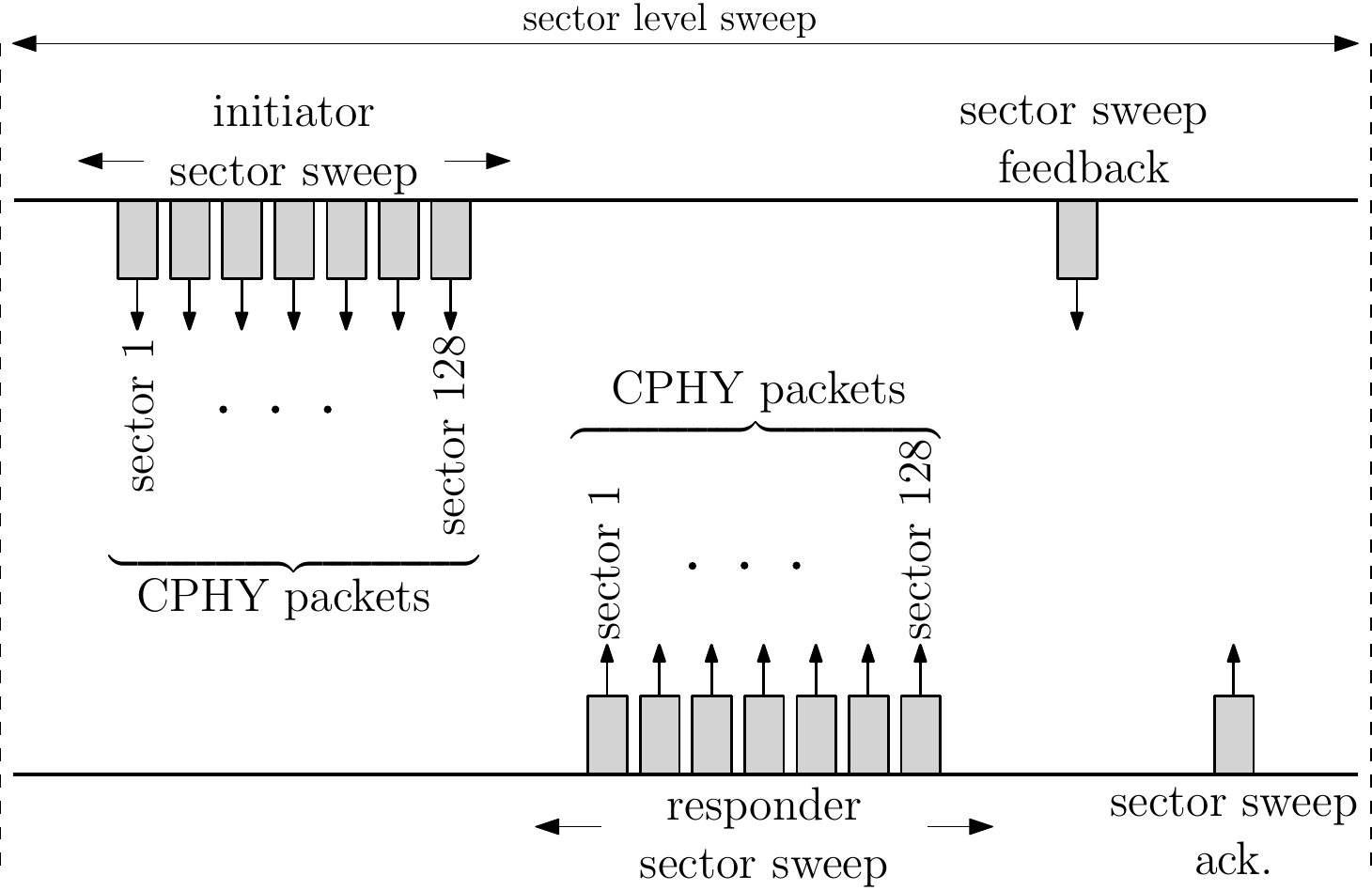}
	\else
	\includegraphics[width=0.50\columnwidth]{fig_01_SLS.pdf}
	\fi	
	\caption{Description of the sector level sweep in the IEEE 802.11ad standard.}
	\label{Fig:SLS}
\end{figure}

To establish a link between two nodes, the IEEE 802.11ad standard executes a beamforming training protocol: the reader may refer to \cite{802.11adStandard,802.11ad-com-mag,Kutty16} and references therein for details. In this paper, we focus on the \gls{sls} described in Fig.~\ref{Fig:SLS}.
In the \gls{iss}, the initiator performs a sequence of directional transmissions towards the available sectors. The responder receives in \gls{qo} mode and determines the transmit sector of the initiator with the largest \gls{snr}. In the \gls{rss}, the role of initiator and responder is reversed. The responder makes a sector sweep, while the initiator receives in \gls{qo} mode and determines the transmit sector of the responder with the largest \gls{snr}: in this phase, each transmission also carries the information as to the best transmit sector for the initiator. Successively, the initiator sends a feedback frame with its best transmit sector and the procedure ends when the responder acknowledges its reception.

\begin{figure}[!tp]
 \centering
  \ifCLASSOPTIONtwocolumn
 \includegraphics[width=0.98\columnwidth]{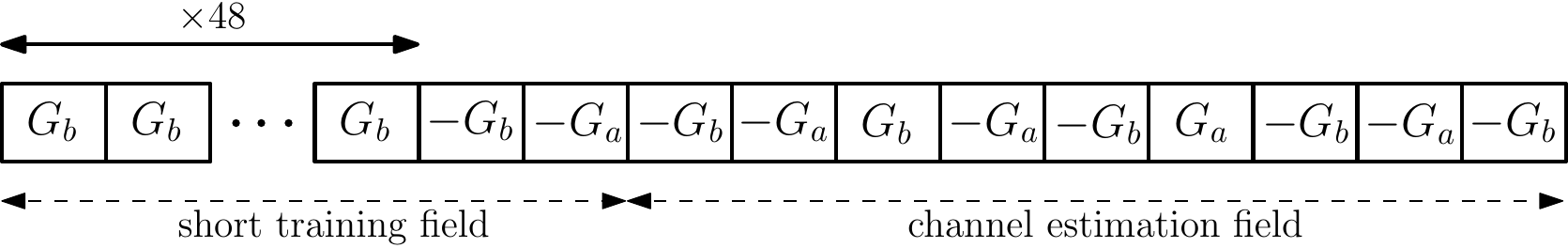}
 \else
 \includegraphics[width=0.60\columnwidth]{fig_02_preamble.pdf}
 \fi
 \caption{Preamble of \gls{cphy} packet in the IEEE 802.11ad standard. $G_{a}$ and $G_{b}$ are Golay complementary sequences of length $128$.}
 \label{Fig:Preamble}
\end{figure}

\begin{figure}[!tp]
 \centering
 \ifCLASSOPTIONtwocolumn
 \includegraphics[width=0.98\columnwidth]{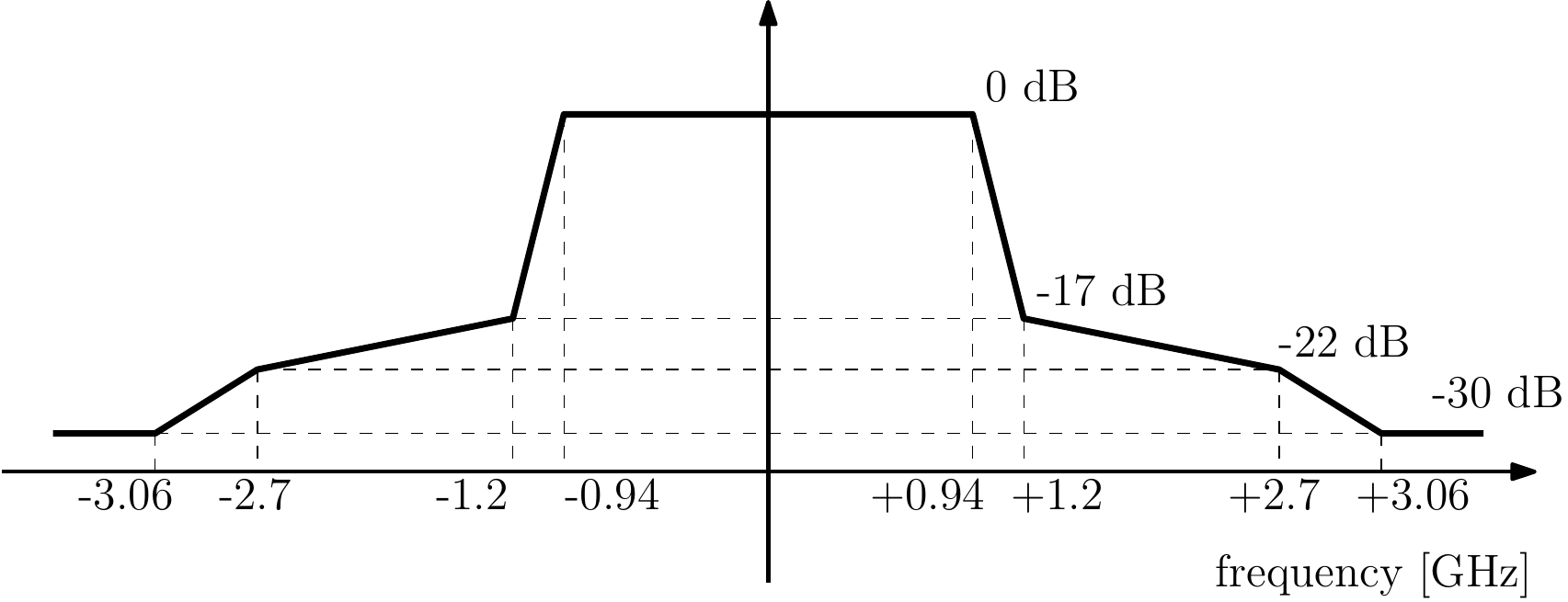}
 \else
 \includegraphics[width=0.50\columnwidth]{fig_03_spectrum_mask.pdf}
 \fi
 \caption{Baseband spectrum mask of the IEEE 802.11ad standard.}
 \label{Fig:spectrum_mask}
\end{figure}

The sequence of directional transmissions performed during the \gls{iss} or \gls{rss} phase can be used for sensing the environment~\cite{Grossi-RADAR-2017,Grossi-VTC-2017,Grossi-TSP-2018}. Each transmission  consists of a \gls{cphy} packet containing a preamble, a header, and a payload~\cite{802.11adStandard,ROHDE_SCHWARZ_2017}.\footnote{An optional beamforming training field can be appended to the \gls{cphy} packet. Without loss of generality, we do not use this optional field.} The preamble is formed by the concatenation of a pair of Golay complementary sequences~\cite{Golay-1961} of length $K_{g}=128$, say $G_a$ and $G_b$, as shown in Fig.~\ref{Fig:Preamble}; the preamble contains $K_{p}=7552$ symbols, grouped in the short training field (STF) and the channel estimation field (CEF).  The header field carries $40$ control bits, while the payload carries a number of informational bits variable from 14 to 1023 bytes. Control and informational bits are scrambled, encoded by using a shortened low-density parity-check code with an effective rate approximatively equal to 1/2, differentially encoded, and spread by a Golay sequence of length $32$, giving an overall packet with $K$ symbols and a throughput of $27.5$ Mbps. The value of $K$ depends on the number of informational bits included in the payload and ranges from $K_{\min}=23168$ to $K_{\max}=539520$. Finally, the digital modulator employs a $\frac{\pi}{2}$-\gls{bpsk} mapping and outputs the following baseband waveform
\begin{equation}
g(t)=\begin{cases}
	\displaystyle \sqrt{{\mathcal P}T}\sum_{k=0}^{K-1} b(k)\psi_{\text{tx}}(t-kT), & t\in[0,T_g] \\
     0, & t\notin[0,T_g]
     \end{cases}
\label{s_expr}
\end{equation}
where  $\{b(k)\}_{k=1}^{K}$ is the sequence of transmitted symbols, $\psi_{\text{tx}}(t)$ is the unit-energy baseband pulse, ${\mathcal P}$ is the transmit power, $T$ is the symbol interval, and $T_g$ ($\geq KT$) is the packet duration (including the latency time between transmissions in successive sectors, if any). The standard specifies the symbol rate, namely, $1/T=1760$ MHz, and the spectrum mask  in Fig.~\ref{Fig:spectrum_mask}, whereby the (one-sided) effective bandwidth is approximatively $W=1/(2T)$; without loss of generality, we assume in the following that $\psi_{\text{tx}}(t)$ has support in $[0, T_{\psi,\text{tx}}]$. 

\subsection{Continuous-time  received signal}
The baseband signal $y(t)\in\mathbb{C}$  received in the sector under inspection is modeled as 
\begin{equation}\label{Eq:RecSignal_1}
	y(t) =	\displaystyle \sum_{p=1}^{P} \alpha_p \e^{2 \pi \i \nu_p t} g(t-\tau_p) + u(t)
\end{equation}
where
\begin{itemize}
	\item $P\in\{0,1,\ldots,P_{\max}\}$ is the unknown number of echoes, with $P_{\max}$ being an upper bound to the number of prospective echoes;
	\item $\alpha_p\in\mathbb{C}$, $\tau_p\in[\tau_{\min},\tau_{\max}]$, and $\nu_p\in[-\nu_{\max},\nu_{\max}]$ are the amplitude, delay, and Doppler shift of the $p$-th echo, for $p=1,\dots,P$, where $\nu_{\max}$, $\tau_{\min}$, and $\tau_{\max}$ are the maximum Doppler shift and the minimum and maximum delay of a prospective echo, respectively; $\alpha_p$ is a function of the two-way antenna gain, the two-way channel response, and the \gls{rcs} of the scatterer causing the reflection;
	
	\item $u(t)\in\mathbb{C}$ is a circularly-symmetric complex Gaussian process, independent of the received echoes, with autocorrelation function $R_{u}(z)=\text{E}[u(t)u^*(t-z)]$, accounting for the thermal noise and, possibly, the residual interference after self-interference cancellation~\cite{Rajagopal_2004,ZhangFD} and weather clutter (if any)~\cite{mmRadarBook}.
\end{itemize}
Prospective echoes come from any target/obstacle present in the environment (for example, another network node, a wall, a person, a tree, a car, etc.); we do not discriminate among scatterers of different nature and any reverberation from the environment, including surface clutter, is deemed as a signal to be detected. We assume that each echo is generated by a distinct and point-like scatterer, deferring to a possible subsequent stage (not included in this study) the task of exploiting some prior knowledge on the inspected area (if any) to associate adjacent detections to  range-spread objects and/or remove ghosts generated by multi-path propagation. 


The signal $y(t)$ is sent to a low-pass causal linear time-invariant filter with impulse response $\psi_{\text{rx}}(t)$ to remove the out-of-bandwidth noise; for example, this can be a filter matched to the transmitted baseband pulse. In the following, we assume that  $\psi_{\text{rx}}(t)$ has support in $[0,T_{\psi,\text{rx}}]$. The filtered signal $r(t)=y(t)\star \psi_{\text{rx}}(t)$ can be written as
\begin{equation}\label{Eq:RecSignal_2}
r(t) \simeq \displaystyle \sum_{p=1}^{P} \alpha_p \e^{2 \pi \i \nu_p t} s(t-\tau_p) + w(t)
\end{equation}
where $s(t)=g(t) \star \psi_{\text{rx}}(t)$ and $w(t)=u(t) \star \psi_{\text{rx}}(t)$ is  a circularly-symmetric complex Gaussian process with autocorrelation function $R_{w}(z)=R_{u}(z)\star \psi_{\text{rx}}(z) \star \psi_{\text{rx}}^*(-z)$. 
The approximation in~\eqref{Eq:RecSignal_2} follows from the fact that $\psi_{\text{rx}}(t)$ has an effective duration of approximatively $1/W$; consequently, we have that
\begin{align}
\e^{2 \pi \i \nu_p t}& g(t-\tau_p)  \star \psi_{\text{rx}}(t)\notag \\
&=\sum_{k=0}^{K-1}b(k) \int_{-\infty}^{+\infty}
\psi_{\text{rx}}(z)\e^{2 \pi \i \nu_p (t-z)}  \notag \\ & \quad \times \psi_{\text{tx}}(t-z-\tau_p-kT)dz \notag\\
&\simeq \sum_{k=0}^{K-1}b(k) \e^{2 \pi \i \nu_p t} \int_{-\infty}^{+\infty}
\psi_{\text{rx}}(z) \psi_{\text{tx}}(t-z-\tau_p-kT)dz \notag\\
&= \e^{2 \pi \i \nu_p t} \left[g(t-\tau_p)  \star \psi_{\text{rx}}(t)\right]
\end{align}
since $\nu_p\ll W$ for all Doppler shifts of interest.
As to this point, recall that $\nu_p=2v_pf_0/c$, where $v_p$ is the radial velocity of the $p$-th object, $f_0$ is the carrier frequency ($60$ GHz), and $c$ is the speed of light; hence, for $v_p=100$ Km/h we have $\nu_p\simeq 1.1\times10^{4}$ Hz and $\nu_p/W\simeq 1.3\times10^{-5}$. 

In the following, we elaborate the portion of $r(t)$ falling in the interval $[T_{w,1},T_{w,2}]\subseteq[0,T_g]$, wherein $T_{w,1}$ and $T_{w,2}$ are the beginning and the end of the processing window, respectively, which are under the designer's control. The end of the processing window cannot exceed $T_g$ since the antenna beam is steered toward the next sector when the transmission of $g(t)$ is concluded. 

\subsection{Range and Doppler accuracy}\label{SEC:Range-Doppler-accuracy}
\begin{figure}[!tp]
	\centering
	\ifCLASSOPTIONtwocolumn
	\includegraphics[width=0.98\columnwidth]{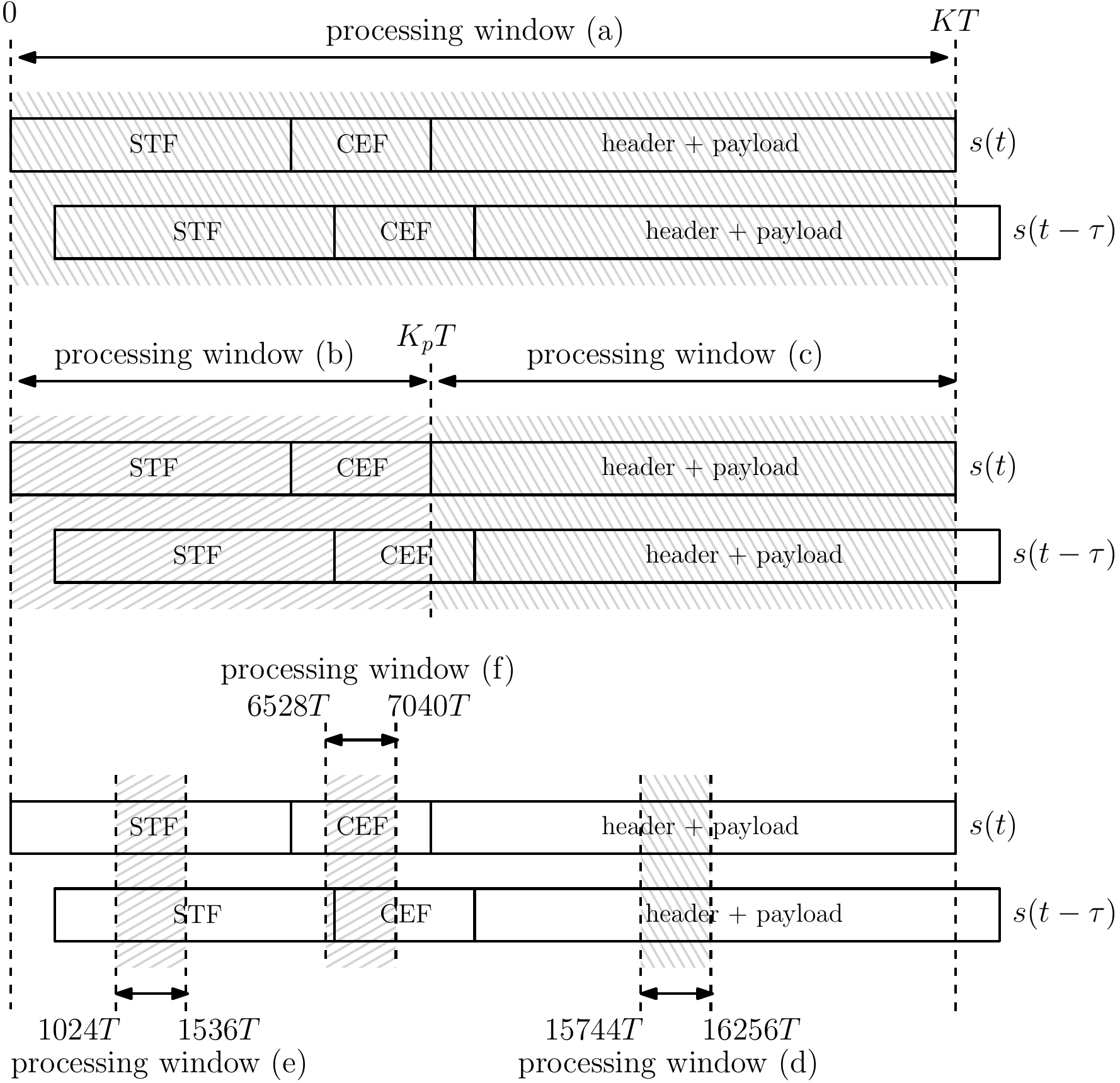}
	\else
	\includegraphics[width=0.60\columnwidth]{fig_04_processing_window.pdf}
	\fi
	\caption{Examples of six processing windows.}
	\label{Fig:processing_window}
\end{figure}

\begin{figure*}[!tp]	
	\subfigtopskip=-2cm
	\centering
	\subfigure[][]{		
		\ifCLASSOPTIONtwocolumn
		\includegraphics[width=0.6\columnwidth]{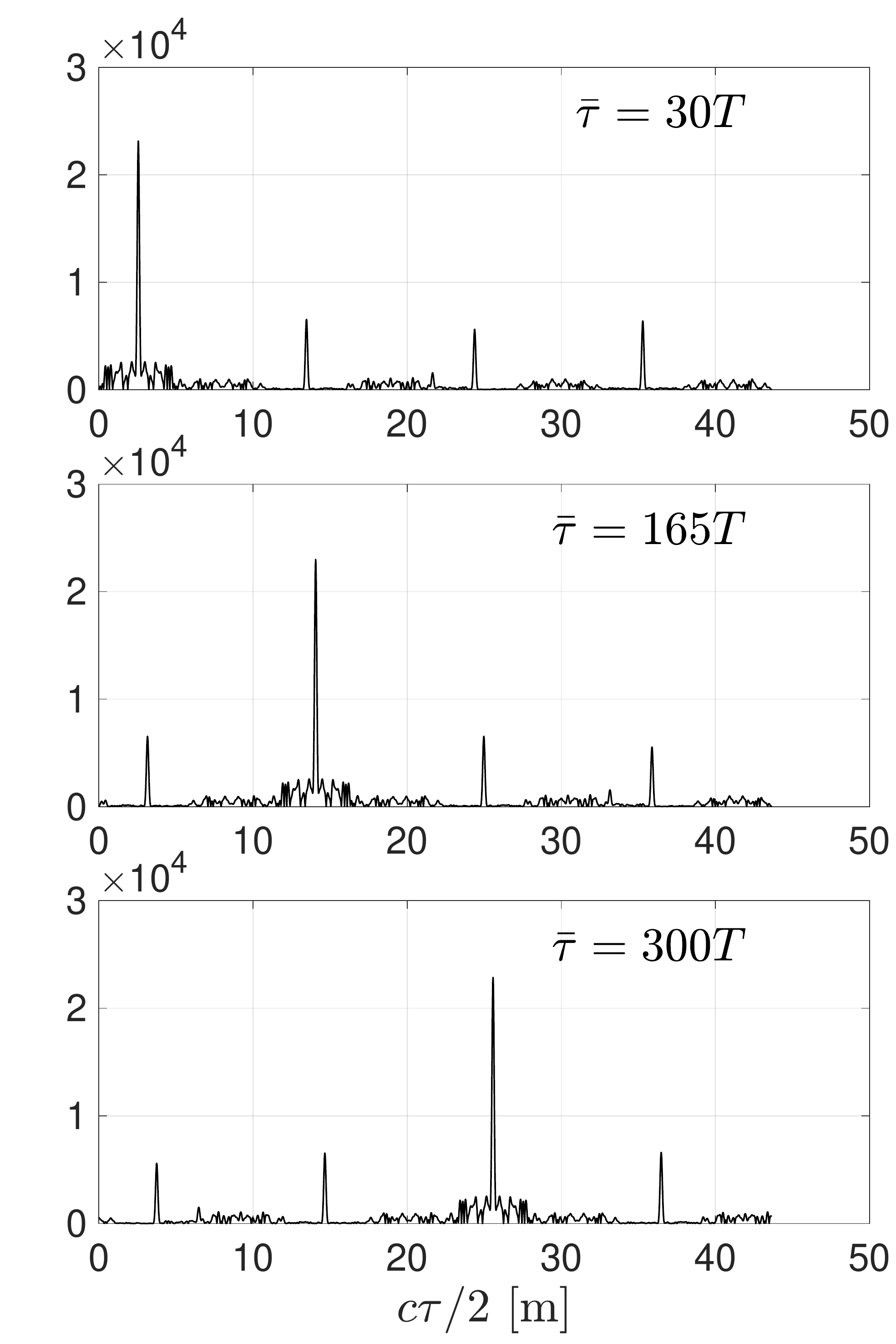}
		\else
		\includegraphics[width=0.30\columnwidth]{fig_05_cor_1.pdf}
		\fi
		\label{fig_cor_1}}
	\subfigure[][]{
		\ifCLASSOPTIONtwocolumn
		\includegraphics[width=0.6\columnwidth]{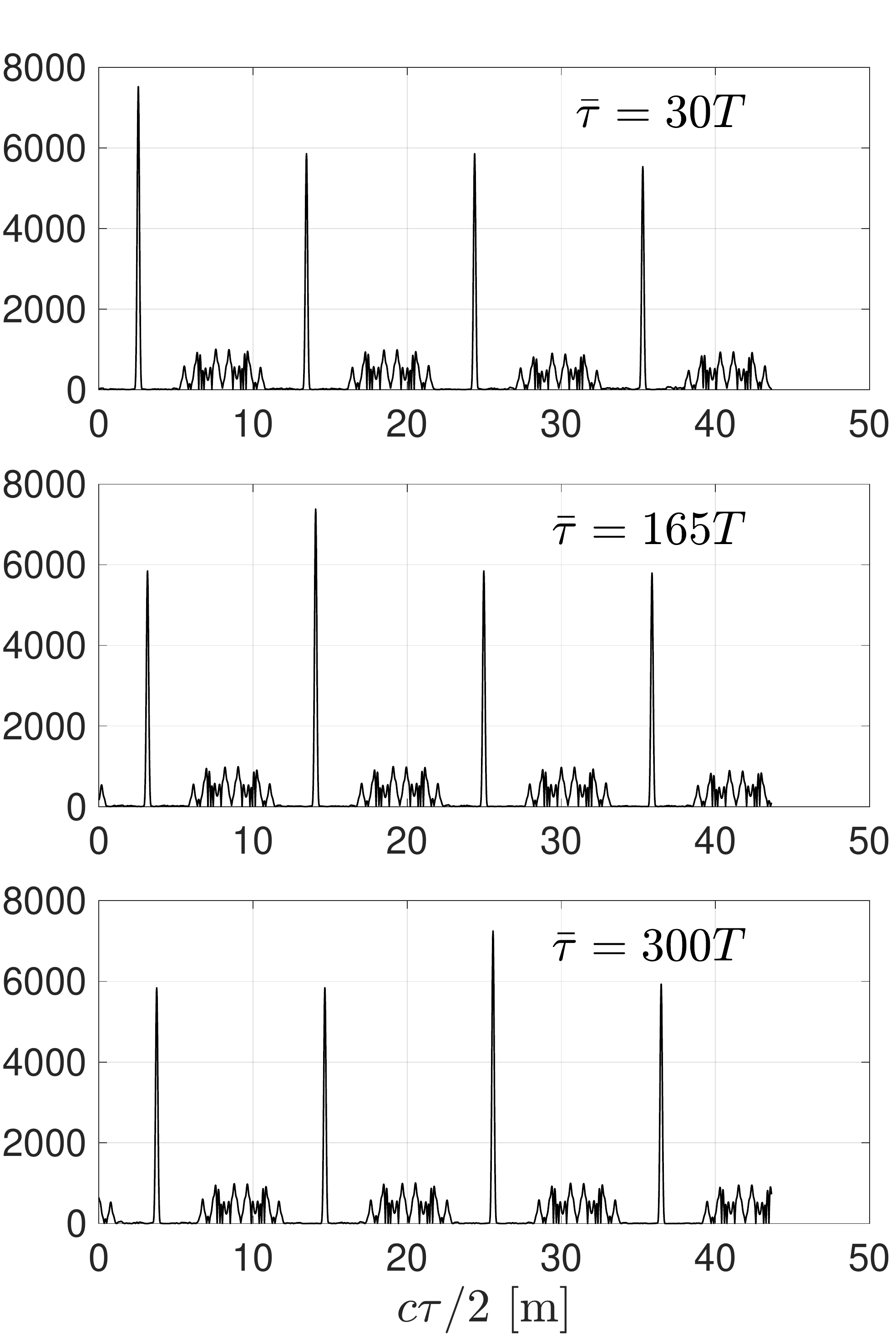}
		\else
		\includegraphics[width=0.30\columnwidth]{fig_05_cor_2.pdf}
		\fi
		\label{fig_cor_2}}
	\subfigure[][]{
		\ifCLASSOPTIONtwocolumn
		\includegraphics[width=0.6\columnwidth]{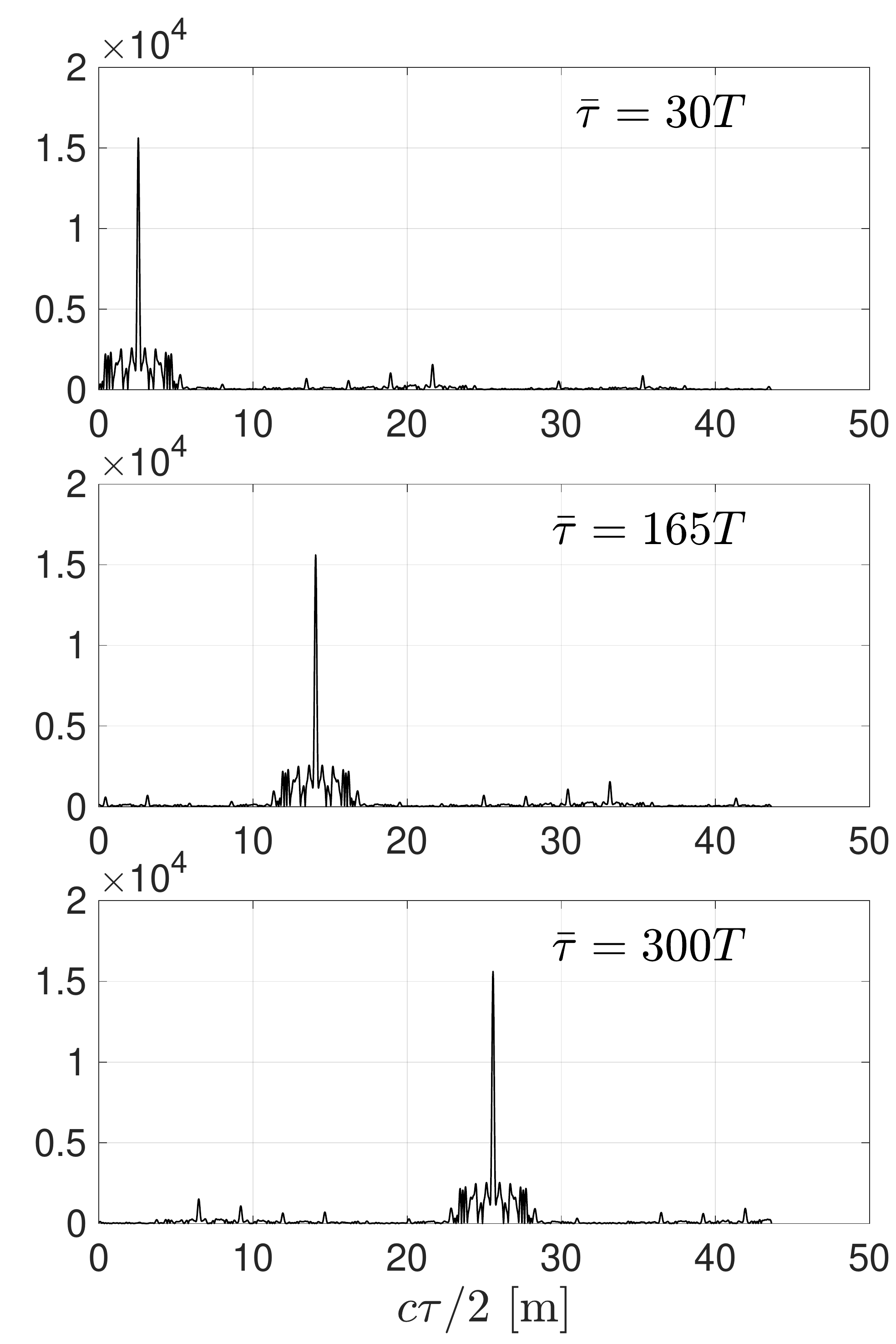}
		\else
		\includegraphics[width=0.30\columnwidth]{fig_05_cor_3.pdf}
		\fi
		\label{fig_cor_3}}
	\\
	\subfigure[][]{
		\ifCLASSOPTIONtwocolumn
		\includegraphics[width=0.6\columnwidth]{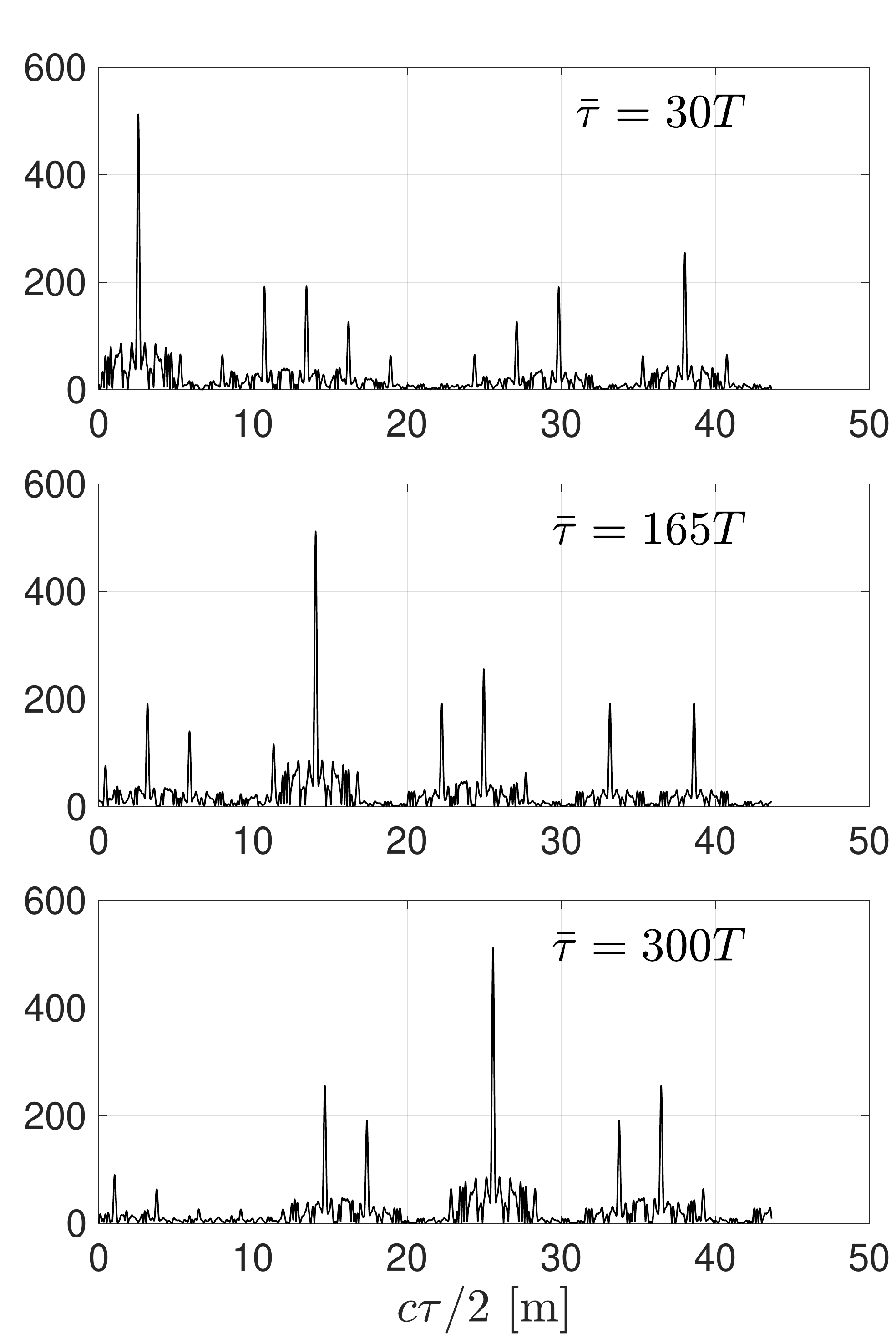}
		\else
		\includegraphics[width=0.30\columnwidth]{fig_05_cor_4.pdf}
		\fi
		\label{fig_cor_4}}
	\subfigure[][]{
		\ifCLASSOPTIONtwocolumn
		\includegraphics[width=0.6\columnwidth]{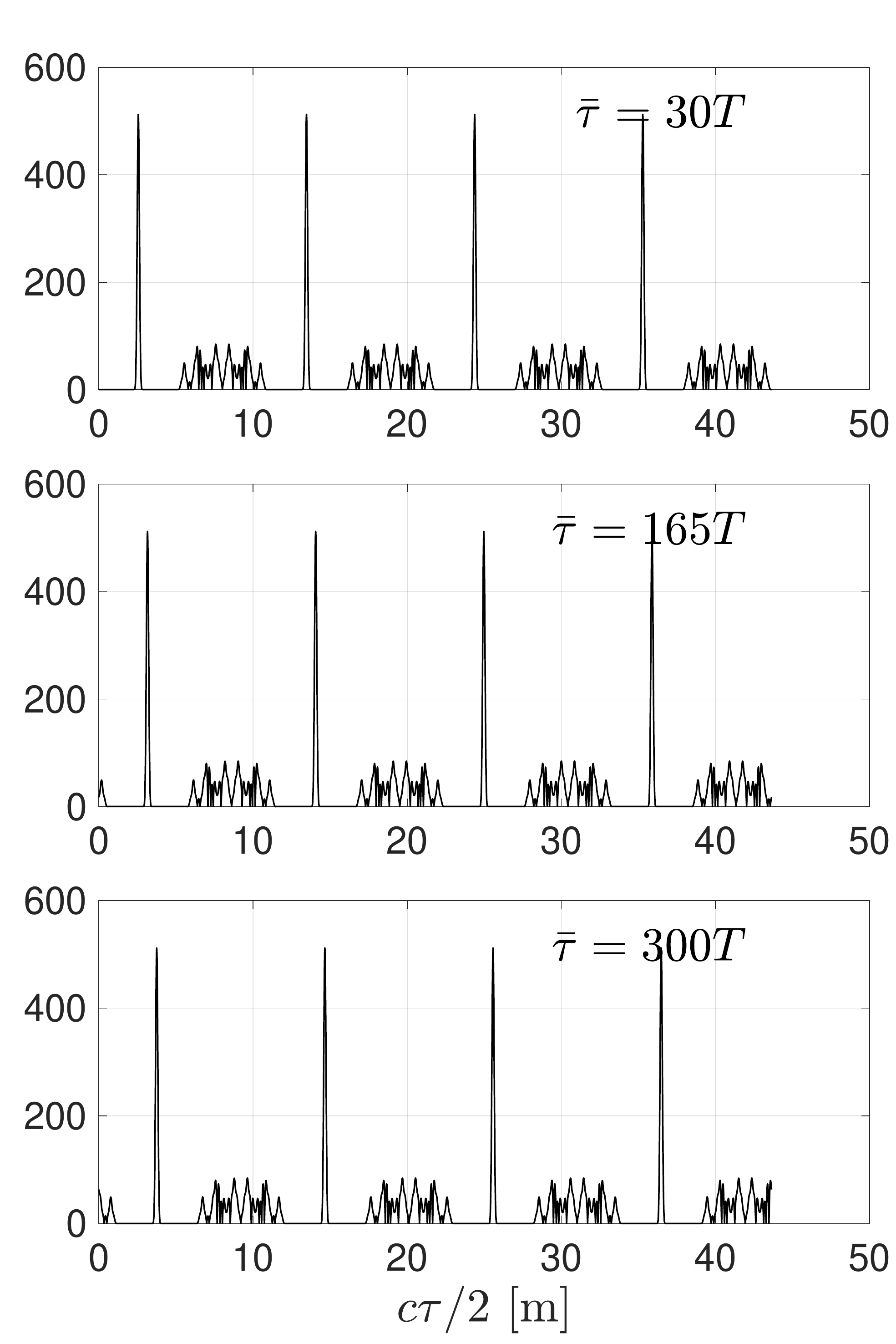}
		\else
		\includegraphics[width=0.30\columnwidth]{fig_05_cor_5.pdf}
		\fi
		\label{fig_cor_5}}
	\subfigure[][]{
		\ifCLASSOPTIONtwocolumn
		\includegraphics[width=0.6\columnwidth]{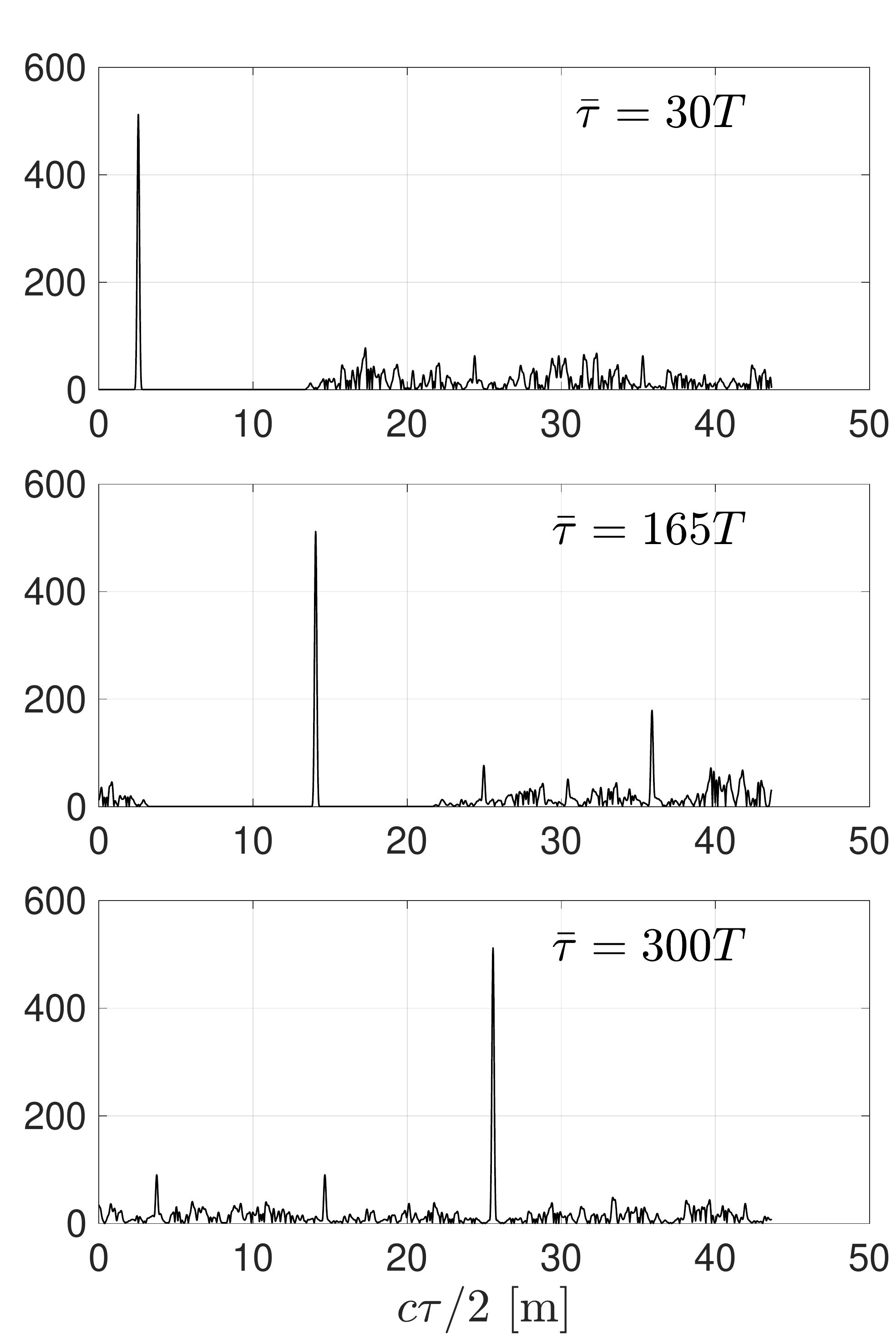}
		\else
		\includegraphics[width=0.30\columnwidth]{fig_05_cor_6.pdf}
		\fi
		\label{fig_cor_6}}
	\caption{ $|\Phi(\tau;\bar{\tau},T_{w,1},T_{w,2})|$ versus $c\tau/2$ for $\bar{\tau}=30T, 165T, 300T$. The processing windows (a)-(f) in Fig.~\ref{Fig:processing_window} are analyzed.}
	\label{fig_cor}
\end{figure*}

\begin{figure}[!tp]
	\centering
	\ifCLASSOPTIONtwocolumn
	\includegraphics[width=0.98\columnwidth]{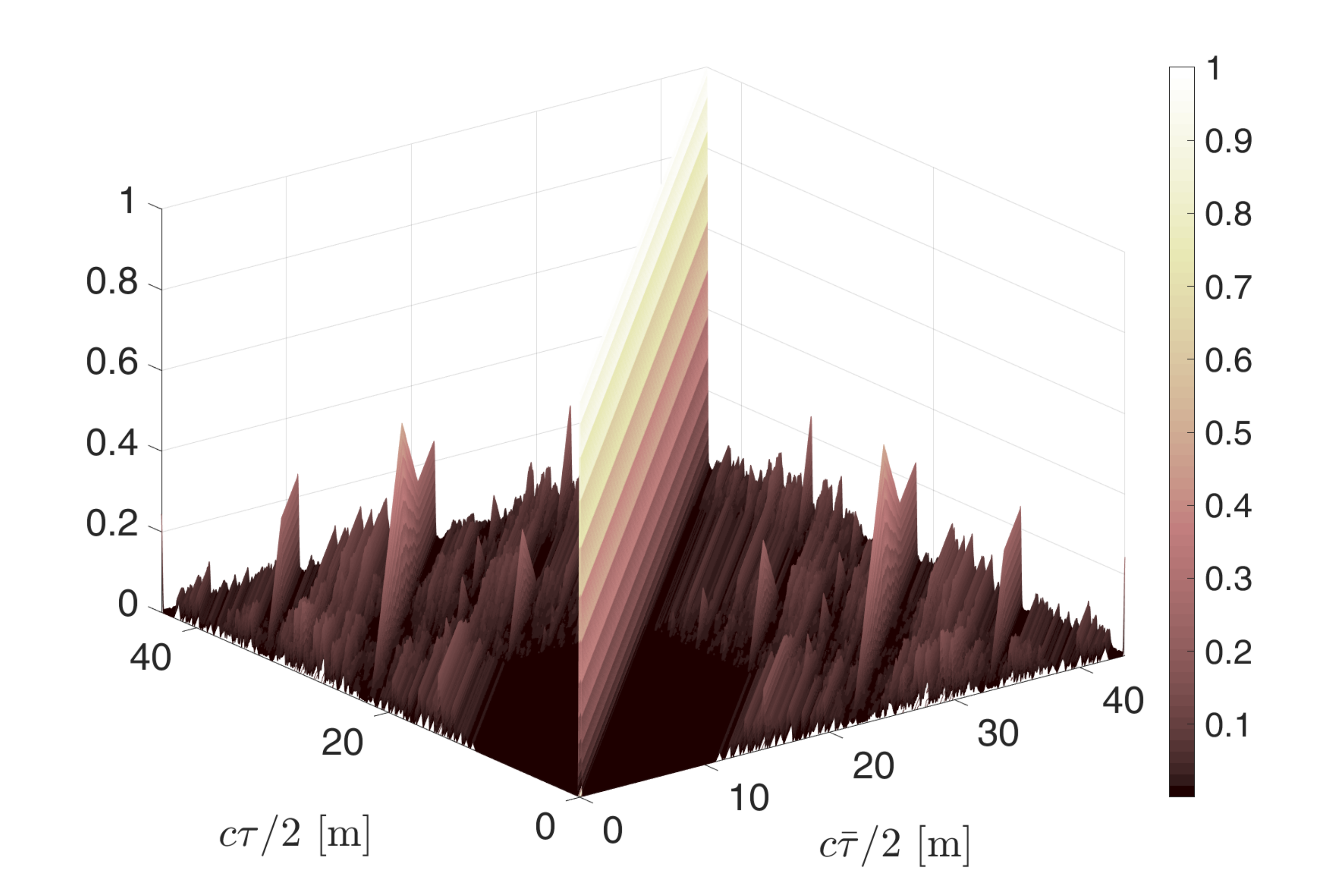}
	\else
	\includegraphics[width=0.60\columnwidth]{fig_06_cor.pdf}
	\fi
	\caption{$\Phi(\tau,\bar{\tau},T_{w,1},T_{w,2})$ (normalized by the height of the main lobe) versus $c\tau/2$ and $c\bar{\tau}/2$ when the processing window (f) is employed.}
	\label{Fig:corr_function_case_f}
\end{figure}


To study the range accuracy, we analyze the similarity between two delayed and truncated (per effect of the limited processing window) copies of the probing signal. To this end, we consider the following windowed correlation function
\begin{align}\label{eq:mod_ambiguity}
&\Phi(\tau,\bar{\tau},T_{w,1},T_{w,2})=\frac{1}{\|\psi(t)\|^2}\int_{T_{w,1}}^{T_{w,2}}s(t-\tau)^{*}s(t-\bar{\tau})dt
\end{align}
where $\|\psi(t)\|^2$ is a normalization factor.
In Fig.~\ref{fig_cor}, we plot $|\Phi|$ versus $c\tau/2$ for $\bar{\tau}=30T, 165T, 300T$ (corresponding to a range of approximately $2.6$, $14.1$, and $25.6$ m, respectively). The inspected values of $\tau$ span $1024$ symbol intervals. We assume that $\psi_{\text{tx}}(t)=\psi_{\text{rx}}(t)$ is a truncated raised cosine pulse with roll-off factor $0.3$ and support in $[0,2T]$ and $K=K_{\min}$; six windows are considered, as shown in Fig.~\ref{Fig:processing_window}, namely
\begin{description}
	\item[(a)] $T_{w,1}=0$ and $T_{w,2}=K T$: this is the largest possible processing interval (recall that $T_g\simeq K T$);
	\item[(b)] $T_{w,1}=0$ and $T_{w,2}=K_{p} T$: this is the time interval during which the preamble is transmitted; 
	\item[(c)] $T_{w,1}=K_{p} T $ and $T_{w,2}=K T$: this is the time interval during which header and payload are transmitted; 
	\item[(d)] $T_{w,1}=15744 T $ and $T_{w,2}=16256 T$: this is a (512 symbol long) portion of the time interval during which header and payload are transmitted; 
	\item[(e)] $T_{w,1}=8 K_{g} T $ and $T_{w,2}=12 K_{g} T$: this is a (512 symbol long) portion of the time interval during which the STF is transmitted; 
	\item[(f)] $T_{w,1}=51 K_{g} T $ and $T_{w,2}=55 K_{g} T$: this is a (512 symbol long) portion of the time interval during which the CEF is transmitted.
\end{description}

The main lobe of $\Phi(\tau,\bar{\tau},T_{w,1},T_{w,2})$ is located at $\tau=\bar{\tau}$ and its full width at half maximum is approximately $1/W$ (corresponding to a range interval of about $17$ cm): this implies that two echoes with the same Doppler shift and delays $\tau_1$ and $\tau_2$, respectively, may not be resolved  if $|\tau_1-\tau_2|< 1/W$ ~\cite{VanTreesIII}. 
Also, as consequence of the normalization adopted in~\eqref{eq:mod_ambiguity}, the height of the main lobe is approximately equal to the number of integrated symbols $(T_{w,2}-\max\{T_{w,1},\bar{\tau}\})/T$ (also referred to as coherent processing gain);\footnote{The approximation stands for the fact that $\psi(t)$ usually embraces more ($4$ in this  example) symbol intervals, thus inducing inter-symbol interference.} for example, in case (a), the height is approximatively equal to $23138$,  $23008$ and  $22868$ for $\bar{\tau}=30T, 165T, 300T$, respectively. 

In cases (a), (b) and (e),  subsidiary peaks occur at multiple of $64cT\approx 10.9$ m, since $48$ repetitions of the Golay sequence $G_b$ are present at the beginning of the \gls{cphy} packet. In case (a), the subsidiary peaks are less pronounced since the processing window is much longer than the preamble duration and, hence, also contains many header and payload symbols for all inspected delays (see also case (c) next). The subsidiary peaks become more pronounced in case (b), since the processing window just contains the STF and a portion of the CEF for all inspected delays, and even indistinguishable for the main peak in case (e), since the processing windows contains a cyclic shift of $4$ repetitions of the Golay sequence $G_b$ for all inspected delays: in the latter two cases, the maximum non-ambiguous range is actually limited to $10.9$ m~\cite{VanTreesIII}.

In case (c), the observed echoes mostly contain header and payload symbols for all inspected delays. This choice provides a well-behaved correlation function, as the encoded control and informational bits form approximatively a pseudo-random sequence. The side lobes around the main peak are caused by the spreading of the encoded bits by a Golay sequence of length 32.  The key point here is that a sufficiently long sequence of encoded bits is elaborated, so that the law of large numbers comes into play. If this processing window is significantly reduced, as in case (d), then the resulting correlation function may get quite poor. 

Case (f) also provides a good correlation function for short ranges. Indeed, the CEF has been specifically designed to take advantage of the complementary property of the Golay sequences $G_a$ and $G_b$ when performing channel estimation~\cite{Liu15}: in particular, the standard exploits the fact that the sum of the aperiodic autocorrelation functions of two complementary sequences is a delta-function~\cite{Golay-1961}. To get a deeper insight on this scenario, we plot in Fig.~\ref{Fig:corr_function_case_f} the windowed correlation function (normalized by its maximum value) versus $c\tau/2$ and $c\bar{\tau}/2$; it is seen that  no  significant subsidiary peak is present if both $c\tau/2$ and $c\bar{\tau}/2$ are less than $21.8$ m. 

As to the Doppler accuracy, notice that, if an echo is received with a delay $\tau$, then the part  falling in the processing window has a duration $T_{w,2}-\max\{T_{w,1},\tau\}$; consequently, two echoes with the same delay $\tau$ and Doppler shifts $\nu_1$ and $\nu_2$, respectively, may not be resolved if $|\nu_1-\nu_2|<\left(T_{w,2}-\max\{T_{w,1},\tau\}\right)^{-1}$~\cite{VanTreesIII}, which in turn amounts to requiring that \begin{equation}
|v_1-v_2|<\frac{c}{2f_0\left(T_{w,2}-\max\{T_{w,1},\tau\}\right)}
\end{equation}
where $v_1$ and $v_2$ are the corresponding radial velocities. Assuming $K=K_{\max}$, $T_{w,1}=0$, and $T_{w,2}=KT$, then the range rate resolution is limited to $30$ Km/h: notice that this value is already unsatisfactory for many  applications. In practice,  the length of the processing window may be  well below $K_{\max}T$ to reduce the complexity~\cite{Grossi-TSP-2018}; for example, for $T_{w,2}-T_{w,1}=K_{p} T$ the resolution is limited to $2000$ Km/h. 

Overall, the above discussion indicates that, while the local accuracy (or resolution) in range is dictated by the signal bandwidth and is in the order of few centimeters, the achievable processing gain and global accuracy (or ambiguity) in range  greatly depends on the adopted processing windows~\cite{VanTreesIII}.  Also, accurate Doppler resolution is not possible in the considered setup due to the short duration of the probing signal, no matter how the processing window is chosen. Finally, we remark that limiting $T_{w,2}$ to $K_{p}T$ ensures that the correlation properties are packet independent.

\subsection{Discrete-time received signal}
The continuous-time signal $r(t)$ is sampled at the time instants $\{T_{w,1}+(m-1)T_c\}_{m=1}^M$, where $M=\left\lfloor (T_{w,2}-T_{w,1})/T_c\right\rfloor+1$ is the number of data samples and $T_c\geq0$ is the sampling interval (which is under the designer's control). After collecting these samples into the vector $\bm{r}=\big(r(T_{w,1}),\ldots,r(T_{w,1}+(M-1)T_c)\big)^T\in\mathbb{C}^M$, we obtain the following discrete-time model
\begin{equation}\label{Eq:RecSignal_3}
	\bm{r} = \begin{cases}
		\displaystyle \bm{X}_{\bm{\tau},\bm{\nu}} \bm{\alpha} + \bm{w}, &\text{ under } \bar{H}_0\\
		\bm{w} ,& \text{ under } H_0
	\end{cases} 
\end{equation}
where
\begin{itemize}
	\item $H_0$ is the hypothesis that no echo is present (i.e., $P=0$) and $\bar{H}_0$ its complement (i.e., $P\in\{1,\ldots,P_{\max}\}$);
	\item $\bm{\alpha}=(\alpha_1,\ldots,\alpha_P)\in \mathbb{C}^P$, $\bm{\tau}=(\tau_1,\ldots,\tau_P)\in [\tau_{\min},\tau_{\max}]^P$, and  $\bm{\nu}=(\nu_1,\ldots,\nu_P)\in [-\nu_{\min},\nu_{\max}]^P$ contain the amplitudes, delays, and Doppler shifts of the received echoes under $\bar{H}_0$;

	\item $\bm{x}_{\tau,\nu}\in\mathbb{C}^{M}$ contains the samples of $\e^{2 \pi \i \nu t} s(t-\tau)$: in the following, we refer to it as the \emph{signature} of an echo with delay $\tau$ and Doppler shift $\nu$; 
	
	\item $\bm{X}_{\bm{\tau},\bm{\nu}}=\big(\bm{x}_{\tau_1,\nu_1},\ldots,\bm{x}_{\tau_P,\nu_P}\big) \in\mathbb{C}^{M\times P}$ contains the signatures of the $P$ echoes under $\bar{H}_0$;
	
	\item $\bm{w}=\big(w(T_{w,1}),\ldots,w(T_{w,1}+(M-1)T_c)\big)^T\in\mathbb{C}^M$ contains the noise samples.
\end{itemize}
Notice that $\bm{w}$ is a circularly-symmetric complex Gaussian vector and the entries of its covariance matrix, say $\bm{C}_{w}$, are $\big[\bm{C}_{w}]_{i,j}=R_{w}\big((i-j)T_c\big)$, for $i=1,\ldots,M$ and $j=1,\ldots,M$. In the following, we assume that $\bm{C}_{w}$ is full-rank. Furthermore, notice that $\bm{x}_{\tau,\nu}=\bm{d}_{\nu}\odot\bm{s}_{\tau}$, where $\odot$ denotes the element-wise Hadamard product,
\begin{equation}\label{signature_vector_d}
\bm{d}_{\nu}=\left(\begin{array}{c}\e^{2 \pi \i \nu (T_{w,1})}\\\e^{2 \pi \i \nu (T_{w,1}+T_c)}\\ \vdots \\ \e^{2 \pi \i \nu \left(T_{w,1}+(M-1)T_c\right)}\end{array}\right)
\end{equation}
is the Doppler steering vector, and 
\begin{equation}\label{signature_vector_s}
\bm{s}_{\tau}=\left(\begin{array}{c}s(T_{w,1}-\tau)\\s(T_{w,1}+T_c-\tau)\\ \vdots \\ s\left(T_{w,1}+(M-1)T_c-\tau\right)\end{array}\right)
\end{equation}
contains the samples of the probing signal  $s(t)$ delayed by $\tau$. If $\tau>T_{w,1}$, then the first $\lfloor(\tau-T_{w,1})/T_c\rfloor+1$ entries of $\bm{s}_{{\tau}}$ are zero. Also, since header and payload change at each transmission, $\bm{s}_{{\tau}}$ is packet-dependent (and, therefore, cannot be computed off-line) for any $\tau$ such that $\tau+K_pT<T_{w,2}$. A sufficient condition to make  $\bm{s}_{{\tau}}$ packet-independent is to set $T_{w,2}\leq K_{p}T$: this may be appealing not only to reduce the implementation complexity of the radar receiver (more on this infra), but also to exploit the correlation properties of the concatenated Golay sequences in the preamble (as explained in Section~\ref{SEC:Range-Doppler-accuracy}).

\section{Detector design}\label{SEC:detector-design}
We are faced here with the problem of detecting an unknown number of echoes and estimating their parameters (namely, amplitude, delay, and  Doppler shift). The work in~\cite{Grossi-TSP-2018} studied this problem  when at most one echo is present, i.e., $P_{\max}=1$. In this case, the log-\gls{lr} is written as~\cite{PoorSignalDetectionBook}
\begin{equation}\label{eq:log-likelihood}
2\Re\left\{\alpha_1\bm{x}_{\tau_1,\nu_1}^{H}\bm{C}_{w}^{-1}\bm{r}\right\}-|\alpha_1|^2\bm{x}_{\tau_1,\nu_1}^{H}\bm{C}_{w}^{-1}\bm{x}_{\tau_1,\nu_1}
\end{equation}
which is maximized over the amplitude $\alpha_1$ when
\begin{equation}\label{beta_est_1}
\alpha_{1}=\frac{\bm{x}_{\tau_{1},\nu_{1}}^H \bm{C}_{w}^{-1}\bm{r}}{ \bm{x}_{\tau_{1},\nu_{1}}^H \bm{C}_{w}^{-1} \bm{x}_{\tau_{1},\nu_{1}}}.
\end{equation}
Hence, the \gls{glrt} becomes 
\begin{equation}\label{eq:GLRT-single-target}
\max_{(\tau,\nu)\in\mathcal{S}} \underbrace{\frac{\left|\bm{x}_{\tau,\nu}^H \bm{C}_{w}^{-1}\bm{r}\right|^2}{\bm{x}_{\tau,\nu}^H \bm{C}_{w}^{-1}\bm{x}_{\tau,\nu}}}_{\mathcal{M}_{\tau,\nu}}\test \gamma 
\end{equation}
where $\mathcal{S}=[\tau_{\min},\tau_{\max}]\times [-\nu_{\max},\nu_{\max}]$ is the feasible search set 
and $\gamma$ is the detection threshold, usually chosen to get a desired probability of false alarm, defined as \begin{equation}P_{\text{fa}}=\Pr(\text{reject }H_0 \text{ under } H_0).\end{equation} Notice that the scoring metric $\mathcal{M}_{\tau,\nu}$ is the squared magnitude of the output of a noise-whitening \gls{mf} normalized by the average noise power at the output of such filter. When a target is detected, the corresponding \gls{ml} estimates of $\tau_1$, $\nu_1$, and $\alpha_1$ are
\begin{align}
(\hat{\tau}_{1},\hat{\nu}_{1})&=\argmax_{(\tau,\nu)\in\mathcal{S}}\mathcal{M}_{\tau,\nu} \\
\hat{\alpha}_{1}&=\frac{\bm{x}_{\hat{\tau}_{1},\hat{\nu}_{1}}^H \bm{C}_{w}^{-1}\bm{r}}{ \bm{x}_{\hat{\tau}_{1},\hat{\nu}_{1}}^H \bm{C}_{w}^{-1} \bm{x}_{\hat{\tau}_{1},\hat{\nu}_{1}}}
\end{align}
respectively. In practice, the maximization over $\mathcal{S}$ is approximated by a discrete search over a uniformly-spaced grid, say \begin{align}\label{eq:grid-definition}\mathcal{G}=\Big\{&(\tau,\nu)\in[J_{\min}\Delta_{g},\ldots,J_{\max}\Delta_{g}]\times [-N \Omega_{g},\ldots,N\Omega_{g}]\Big\}\end{align} where $\Delta_{g}$ and $\Omega_{g}$ are the quantization step sizes in the delay and Doppler domain, respectively,  $J_{\min}=\lceil \tau_\text{min}/\Delta_{g}\rceil$, $J_{\max}=\lfloor \tau_\text{max}/\Delta_{g}\rfloor$, and $N=\lfloor \nu_\text{max}/\Omega_{g}\rfloor$~\cite{VanTreesIII}. Notice that, if  $2v_{\max}f_0/c\ll \left(T_{w,2}-\max\{T_{w,1},\tau_{\min}\}\right)^{-1}$, then  $\bm{d}_{\nu}\simeq \bm{1}_{M}$ for all inspected Doppler shifts and, hence, the Doppler search can be avoided (i.e., $N=0$); for example, for $v_{\max}=100$ Km/h, we need $\left(T_{w,2}-\max\{T_{w,1},\tau_{\min}\}\right) \ll 90 $ $\mu$s.

When $P_{\max}>1$, a natural adaptation of the above procedure is to declare the presence of a target at any point $(\tau,\nu)\in\mathcal{G}$ where $\mathcal{M}_{\tau,\nu}$ exceeds $\gamma$. 
This \gls{mf}-based procedure would be optimal (in the GLRT sense~\cite{Stuller-1975}) if the whitened signatures $\{\bm{C}_{w}^{-1/2}\bm{x}_{\tau,\nu}\}_{(\tau,\nu)\in\mathcal{G}}$ corresponding to the possible look directions were orthogonal. Since this condition is not verified here, a target located at $(\bar{\tau},\bar{\nu})$ produces a signal spillover at a different look direction  $(\tau,\nu)\neq(\bar{\tau},\bar{\nu})$, possibly causing spurious threshold crossings (that are false detections) and the masking of weaker targets\footnote{This problem also arises in code-division multiple-access (CDMA) communication systems~\cite{Verdu-1998}, where it is referred to as near-far problem.}~\cite{Blunt-Gerlach-2006,Kulpa-2013}. A peak-finding algorithm can be employed after the thresholding operation, thus treating all detections in a small neighborhood of each peak as false detections due to a \emph{local} signal spillover:  this strategy, referred to as \gls{mf-pd}, cannot however recognize false detections generated by the signal spillover at positions much distant from a true scattering point neither it can discriminate the echo of a target actually present at a given look direction from the interference caused by a target located elsewhere. 



\subsection{IIC-AMF}
To overcome the limitations of the \gls{mf-pd}, for each look direction, the receive filter should be closely matched to the desired echo while also suppressing (part of) the interference caused by other echoes~\cite{Kelly-1992,Kelly-1986,Li-Stoica-2005}. To achieve this goal, we propose an iterative adaptive procedure which attempts to extract and detected the prospective echoes one-by-one from the observed signal, after removing the interference from the previously detected (stronger) targets. 

To be more specific, assume that $p-1$ echoes have been detected in the previous $p-1$ steps and let  $\{\hat{\alpha}_n,\hat{\tau}_n,\hat{\nu}_n\}_{n=1}^{p-1}$ be the corresponding estimates of the amplitudes, delays, and Doppler shifts.
At the next iteration, we consider the problem of detecting a prospective echo with amplitude $\alpha_p$, delay $\tau_p$, and Doppler shift $\nu_p$ in the presence of  $p-1$ interfering signals and noise. The proposed detector operates by maximizing over  $(\tau,\nu)$ the following data-adaptive metric
\begin{equation}\label{eq:IIC-AMF-metric}
\mathcal{M}_{\tau,\nu}(p)=\frac{\left|\bm{x}_{\tau,\nu}^H \left(\bm{C}_{i}(p)+\bm{C}_{w}\right)^{-1}\bm{r}\right|^2}{\bm{x}_{\tau,\nu}^H \left(\bm{C}_{i}(p)+\bm{C}_{w}\right)^{-1}\bm{x}_{\tau,\nu}}
\end{equation}
where $\bm{C}_{i}(p)$ represents the interference covariance matrix at iteration $p$, for $p=1,\ldots,P_{\max}$.  Notice that~\eqref{eq:IIC-AMF-metric} can be interpreted as the instantaneous power at the output of a noise-plus-interference whitening matched filter normalized by the average interference-plus-noise power at the output of such filter, whereby resembling the detector proposed  in~\cite{Kelly-1992}.  

Since a set of secondary data is not available here, we resort to a parametric estimate of $\bm{C}_{i}(p)$. At the design stage, we assume that the $n$-th interfering signal comes from a target whose delay and Doppler shift, say $(\tau_n,\nu_n)$, lay anywhere in the region $(\hat{\tau}_n-E_n,\hat{\tau}_n+E_n)\times[\hat{\nu}_n-\Theta_n,\hat{\nu}_n+\Theta_n]\subset\mathcal{S}$, for $n=1,\ldots,p-1$, and that the  corresponding amplitude is $|\hat{\alpha}_n|e^{-\i \hat{\phi}_n}$, where $\{\hat{\phi}_n\}_{n=1}^{p-1}$ are independent random phases uniformly-distributed over $[0,2\pi]$; then, the interference covariance matrix is computed as follows
\begin{equation}\label{eq:covariance_matrix}
\bm{C}_{i}(p)=\begin{cases}\bm{0}, & \text{if } p=1\\ 
\displaystyle \sum_{n=1}^{p-1}|\hat{\alpha}_n|^2\bm{Q}_n,
& \text{if } p\in\{2,3,\ldots,P_{\max}\}
\end{cases}
\end{equation}
where
\begin{equation}
\bm{Q}_n=\text{E}\left[\bm{x}_{\hat{\tau}_n+\epsilon_n,\hat{\nu}_n+\theta_n}\bm{x}_{\hat{\tau}_n+\epsilon_n,\hat{\nu}_n+\theta_n}^H\right]
\end{equation}
and $\epsilon_n$ and $\theta_n$ are independent random variables with a uniform distribution in $[-E_n,E_n]$ and $[-\Theta_n,\Theta_n]$, respectively.  $\{E_n,\Theta_n\}_{n=1}^{p-1}$ are positive design parameters accounting for the inherent localization errors at the previous steps of the algorithm: the intuition here is that the signature $\bm{x}_{\tau_n,\nu_n}$ must be contained in the column span of $\bm{Q}_n$ whenever \emph{local} estimation errors are made. Leveraging the analysis in Section~\ref{SEC:description} and  the Cramér-Rao Bound on the variance of any unbiased estimator of the unknown target parameters derived in~\cite{Grossi-TSP-2018}, we set
\begin{subequations}
	\begin{align}\label{eq:E_THETA}
	E_n&=\frac{1}{2}\max\left\{\Delta_g, \lambda_n^{-1/2}   \left(2\pi W\right)^{-1}\right\}\\
	\Theta_n&=\frac{1}{2}\max\left\{\Omega_g, \lambda_n^{-1/2} \left(T_{w,2}-\max\{T_{w,1},\tau_{\max}\}\right)^{-1} \right\}
	\end{align}
\end{subequations}
where $\lambda_n$ is a tuning parameter proportional to $\mathcal{M}_{\hat{\tau}_n,\hat{\nu}_n}(n)$, in keeping with the fact that a smaller estimation error occurs if the (estimated) \gls{sinr} is larger.  Since a detection is declared only if $\mathcal{M}_{\hat{\tau}_n,\hat{\nu}_n}(n)>\gamma$, a conservative (and data-independent) choice is to set $\lambda_n$  proportional to the  threshold.

If the maximum of $\mathcal{M}_{\tau,\nu}(p)$ exceeds the threshold $\gamma$, a target detection is declared; also, the corresponding estimates of the delay and Doppler shift, say $(\hat{\tau}_p,\hat{\nu}_p)$, are readily computed from the location of the maximum, while an estimate of the target amplitude is
\begin{equation}\label{eq:IIC-AMF-beta}
\hat{\alpha}_{p}=\frac{\bm{x}_{\hat{\tau}_{p},\hat{\nu}_{p}}^H \left(\bm{C}_{i}(p)+\bm{C}_{w}\right)^{-1}\bm{r}}{ \bm{x}_{\hat{\tau}_{p},\hat{\nu}_{p}}^H \left(\bm{C}_{i}(p)+\bm{C}_{w}\right)^{-1} \bm{x}_{\hat{\tau}_{p},\hat{\nu}_{p}}}.
\end{equation}
Otherwise, the search is terminated. We refer to this detector, summarized in Algorithm~\ref{ALG-IIC-AMF}, as the \gls{amfd} with \gls{iic}, shortly IIC-AMFD.

\begin{algorithm}[!pt]
	\caption{IIC-AMF detector}
	\begin{algorithmic}[1]\label{ALG-IIC-AMF}
		\STATE Set $\gamma$ based on the desired $P_{\text{fa}}=\Pr(\text{reject }H_0 \text{ under } H_0)$ 
		\STATE Set $\mathcal{G}_{1}=\mathcal{G}$ and  $\hat{P}=0$	
		\FOR{$p=1,\ldots,P_{\max}$}	
		\STATE Compute $\displaystyle (\hat{\tau},\hat{\nu})=\arg\max_{(\tau,\nu)\in\mathcal{G}_p} \mathcal{M}_{\tau,\nu}(p)$
		\IF{$\mathcal{M}_{\hat{\tau},\hat{\nu}}(p)>\gamma$}	
		\STATE $\hat{P}=\hat{P}+1$ 
		\STATE Set $(\hat{\tau}_p,\hat{\nu}_p)=\left(\hat{\tau},\hat{\nu}\right)$ and compute $\hat{\alpha}_p$ from~\eqref{eq:IIC-AMF-beta}
		\STATE Update the search set \begin{align}\mathcal{G}_{p+1}=\mathcal{G}_p\setminus\big\{&(\tau,\nu)\in\mathcal{G}_p:\,|\tau-\hat{\tau}_p|\leq E_p \notag \\ &\text{ and } |\nu-\hat{\nu}_p|\leq \Theta_p\big\}\notag\end{align}
		\ELSE 
		\STATE \bf{break}
		\ENDIF
		\ENDFOR
		\STATE The number of detected targets is $\hat{P}$; if $\hat{P}\geq1$, the estimated amplitudes, delays, and Doppler shifts are
		$\left\{(\hat{\alpha}_p,\hat{\tau}_p,\hat{\nu}_p)\right\}_{p=1}^{\hat{P}}$
	\end{algorithmic}
\end{algorithm}

\subsubsection{Implementation complexity} Each iteration of Algorithm~\ref{ALG-IIC-AMF} (and therefore the detection of each target) mainly requires computing the inverse of the matrix $\bm{C}(p)=\bm{C}_{i}(p)+\bm{C}_{w}$ and, for all grid points, the pair of vector-matrix-vector products $\bm{x}_{\tau,\nu}^H\left(\bm{C}(p)\right)^{-1}\bm{r}$ and $\bm{x}_{\tau,\nu}^H\left(\bm{C}(p)\right)^{-1}\bm{x}_{\tau,\nu}$, which have a computational complexity $\mathcal{O}\left(M^3\right)$ and $\mathcal{O}\left(|\mathcal{G}|M^2\right)$, respectively. If $T_{w,2}\leq K_{p}T$ and $\lambda_n$ is data-independent, the matrix $\bm{Q}_n$ can be precomputed off-line for any $(\hat{\tau}_n\,\hat{\nu}_n)\in\mathcal{G}$, thus simplifying the detector implementation. In particular, let $\bm{U}_n\hat{\bm{\Lambda}}_n\bm{U}_n^{H}$ be the economy-size singular value decomposition of $\bm{Q}_n$, where $\bm{U}_n\in\mathbb{C}^{M\times d_n}$ is a matrix with orthogonal columns, $\bm{\Lambda}_n$ is a $d_n\times d_n$ diagonal matrix with diagonal entries $\lambda_{n,1}\geq\cdots\geq \lambda_{n,d_n}>0$, and $d_n$ is the rank of $\bm{Q}_n$. Then, we can leverage the matrix inversion lemma to iteratively update the inverse of $\bm{C}(p)$ for $p=2,\ldots,P_{\max}$, i.e.,
\begin{align}
\left(\bm{C}(p)\right)^{-1}&=\left(\bm{C}(p-1)\right)^{-1}- \left(\bm{C}(p-1)\right)^{-1}\bm{U}_p 
\notag \\ & \quad \times
\left(\bm{\Lambda}_p^{-1}+\bm{U}_p^{H}\left(\bm{C}(p-1)\right)^{-1} \bm{U}_p\right)^{-1}  \notag\\ & \quad  \times
\bm{U}_p^{H}\left(\bm{C}(p-1)\right)^{-1}.
\end{align}

\subsubsection{Refined parameter estimation}\label{Sec:refined-estimate}
Assume that $\hat{P}\geq1$ targets have been detected and, for $p=1,\ldots,\hat{P}$, consider the following  scoring metric 
\begin{equation}\label{eq:IIC-AMF-metric-modified}
	\bar{\mathcal{M}}_{\tau,\nu}(p)=\frac{\left|\bm{x}_{\tau,\nu}^H \left(\bar{\bm{C}}_{i}(p)+\bm{C}_{w}\right)^{-1}\bm{r}\right|^2}{\bm{x}_{\tau,\nu}^H \left(\bar{\bm{C}}_{i}(p)+\bm{C}_{w}\right)^{-1}\bm{x}_{\tau,\nu}}
\end{equation}
with
\begin{equation} \bar{\bm{C}}_{i}(p)= \sum_{\substack{ n=1 \\ n\neq p}}^{\hat{P}}|\hat{\alpha}_n|^2\bm{Q}_n
\end{equation}
which differs from $\mathcal{M}_{\tau,\nu}(p)$ in~\eqref{eq:IIC-AMF-metric} for the fact that all detected signals, expect the $p$-th signal, are now included in the construction of the interference covariance matrix. A refined estimate of the delay and Doppler shift of the $p$-th target is now obtained as follows  
\begin{equation}\label{eq:refined_localization}
(\bar{\tau}_p,\bar{\nu}_p)=\arg\max_{(\tau,\nu)\in \mathcal{B}_p}	\bar{\mathcal{M}}_{\tau,\nu}(p)
\end{equation}
where $\mathcal{B}_p$ is a small search region around the initial estimate $(\hat{\tau}_p,\hat{\nu}_p)$, namely,
\begin{equation}
	\mathcal{B}_p=\big\{(\tau,\nu)\in\mathcal{S}:\,|\tau-\hat{\tau}_p|\leq E_p \text{ and } |\nu-\hat{\nu}_p|\leq \Theta_p\big\}.
\end{equation}
The local maximization in~\eqref{eq:refined_localization} can be implemented without much computational effort by using a fine-grid search in $\mathcal{B}_p$. 
Accordingly, we can also compute a refined estimate of the amplitude of the target as follows
\begin{equation}\label{eq:IIC-AMF-beta-refined}
	\bar{\alpha}_{p}=\frac{\bm{x}_{\bar{\tau}_{p},\bar{\nu}_{p}}^H \left(\bar{\bm{C}}_{i}(p)+\bm{C}_{w}\right)^{-1}\bm{r}}{ \bm{x}_{\bar{\tau}_{p},\bar{\nu}_{p}}^H \left(\bar{\bm{C}}_{i}(p)+\bm{C}_{w}\right)^{-1} \bm{x}_{\bar{\tau}_{p},\bar{\nu}_{p}}}.
\end{equation}

\section{Numerical results}\label{SEC:numerical-results}
\begin{figure}[!tp]
	\centering
	\ifCLASSOPTIONtwocolumn
	\includegraphics[width=0.98\columnwidth]{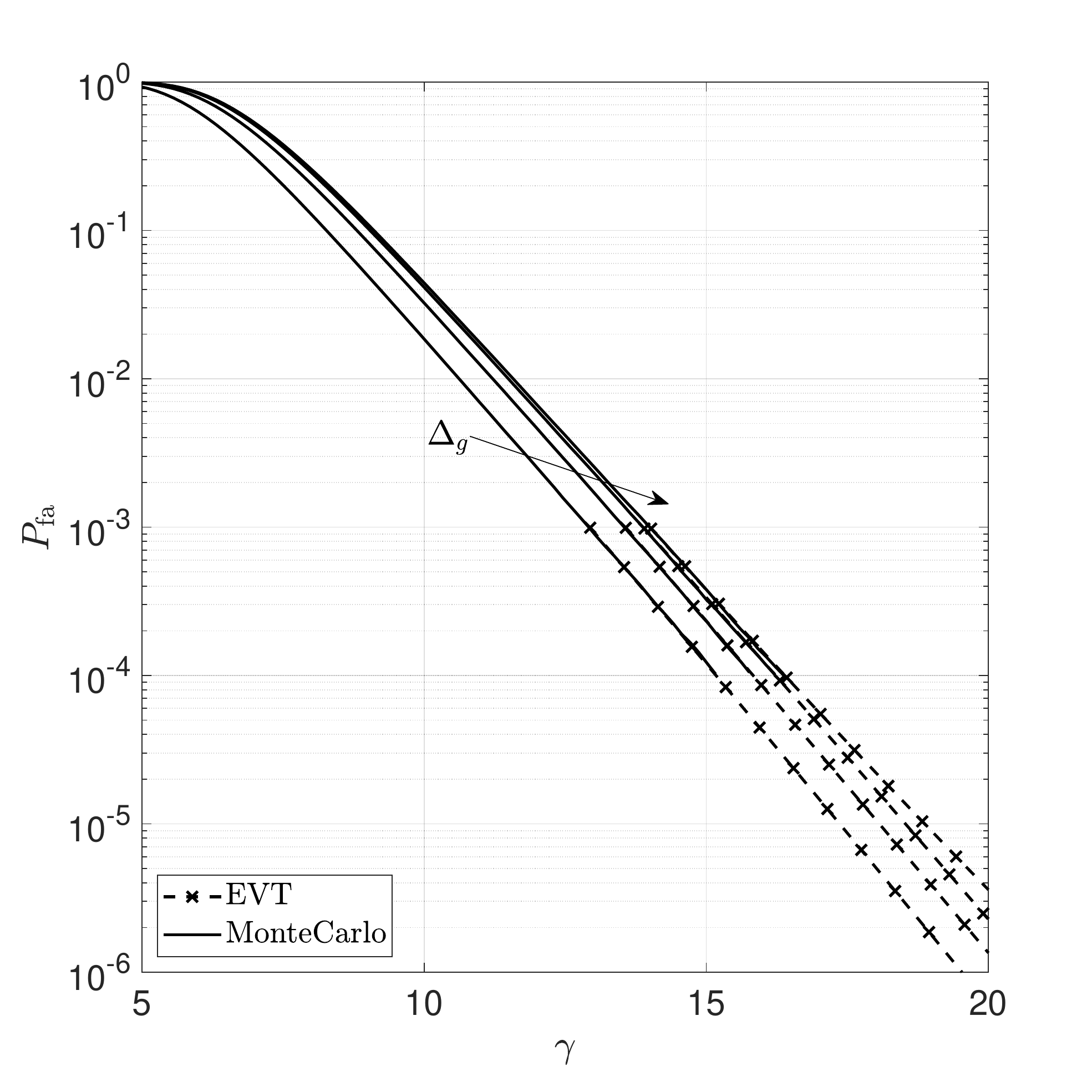}
	\else
	\includegraphics[width=0.50\columnwidth]{fig_8}
	\fi
	\caption{$P_{\text{fa}}$ versus $\gamma$ for $\Delta_g=T,T/2,T/4,T/8$. $\Delta_g$ decreases in the direction of the arrow.}
	\label{fig_pfa_1}
\end{figure}

We present here some numerical examples to asses the performance of the proposed detector. We consider a network with a transmit power of $\mathcal{P}=10$~mW; we assume a thermal noise with a flat power spectral density of $\sigma_{u}^{2}=-177$~dBm/Hz and a noise figure of the front-end receiver of $F_{u}=7$~dB, whereby the autocorrelation of $u(t)$ in~\eqref{Eq:RecSignal_1} becomes $R_{u}(z)=F_{u}\sigma_{u}^2 \delta(z)$. We model the amplitude of the $p$-th target as
\begin{equation}\alpha_p=\sqrt{G A_p^{\text{slow}}A_p^{\text{fast}}\zeta_p/L_p}e^{-j\phi_p}\end{equation}
where $G=46$ dBi is the two-way (i.e., transmit and receive) antenna gain,  $10\log_{10}(A_p^{\text{slow}})$ is  a Gaussian random variable with a zero-mean and a standard deviation of $3$ dB (accounting for log-Normal shadowing), $(A_p^{\text{fast}})^{1/2}$ is a Rice random variable with a unit-power and a shape parameter of $15$ dB (accounting for fast fading), $\zeta_p$ is the \gls{rcs} of the target, $L_p=(4\pi)^3\lambda^{-2}\left(c\tau_p/2\right)^{4}$ is the two-way path loss (computed according to the radar equation), and $\phi_p$ is a random phase uniformly distributed over $[0,2\pi)$~\cite{Rappaport_2015,ChannelModelIII,SkolnikBook}. Also, we assume that $\psi_{\text{tx}}(t)=\psi_{\text{rx}}(t)$ is a truncated raised cosine pulse with roll-off factor $0.3$ and support in $[0,2T]$, which satisfies the spectrum mask in~\eqref{Fig:spectrum_mask}, and that $K=K_{\min}$. 

In the following, we set the sampling rate at twice the signal bandwidth $W$, whereby $T_c=T$, and adopt the processing window (f) of Figure~\ref{Fig:processing_window}, which provides reasonably good correlation properties---see Figure~\ref{fig_cor_6}---at an affordable cost, as only $M=513$ samples are processed; also, we consider a minimum and maximum range of $r_{\min}=c\tau_{\min}/2=5$~m and $r_{\max}=c\tau_{\max}/2=40$~m, respectively, a maximum speed of $v_{\max}=c\nu_{\max}/(2 f_0)=5$~m/s, and $P_{\max}=\lfloor(\tau_{\max}-\tau_{\min})/T\rfloor$.
In the implementation of proposed detector, we set  $\lambda_n=\mathcal{M}_{\hat{\tau}_n,\hat{\nu}_n}(n)/16$ in~\eqref{eq:E_THETA}; also, we set $N=0$ in~\eqref{eq:grid-definition}, since Doppler search is unnecessary (see the discussion in Section~\ref{SEC:Range-Doppler-accuracy}), and consider four values for $\Delta_{g}$, namely, $T$, $T/2$, $T/4$, and $T/8$. Finally, the refined estimates in Section~\ref{Sec:refined-estimate} are computed by using a delay step size of $T/512$. The performance of the IIC-AMFD is compared with that of the MF-PD and with that of the \gls{glrt} in~\eqref{eq:GLRT-single-target} when at most one target is present in the scene (i.e., $P_{\max}=1$); we refer to this latter solution as the single-target detector (STD). 

\begin{figure}[!tp]
	\centering
	\ifCLASSOPTIONtwocolumn
	\includegraphics[width=0.98\columnwidth]{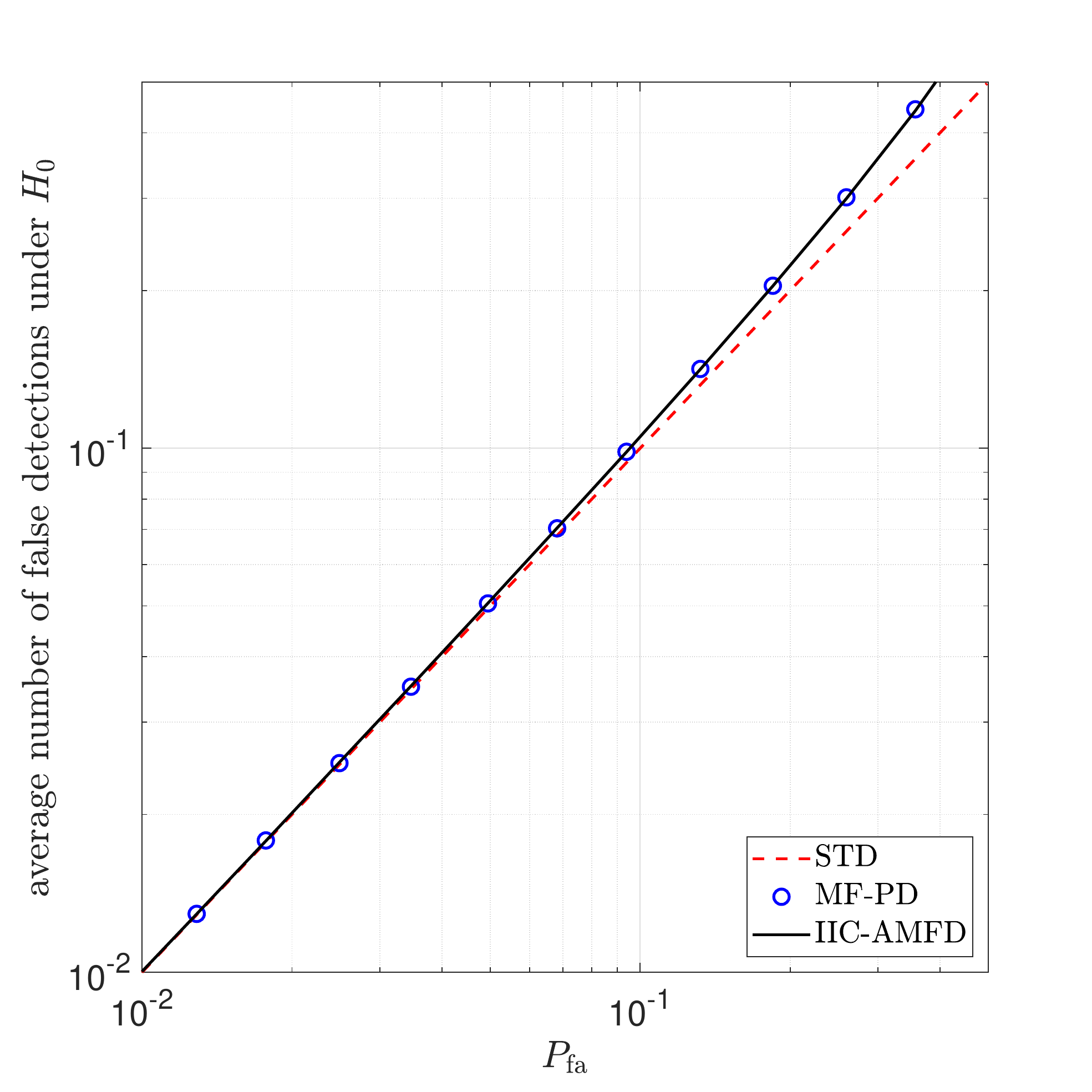}
	\else
	\includegraphics[width=0.50\columnwidth]{fig_9}
	\fi
	\caption{Average number of false detections declared under $H_0$ versus $P_{\text{fa}}$ when $\Delta_g=T$.}
	\label{fig_pfa_2}
\end{figure}

We first investigate the threshold setting. In Figure~\ref{fig_pfa_1}, we plot $P_{\text{fa}}$ versus $\gamma$ for different values of $\Delta_g$. For all considered schemes (namely, STD, MF-PD, IIC-AMD), a false alarm event occurs if\footnote{Notice that $\mathcal{M}_{\tau,\nu}=\mathcal{M}_{\tau,\nu}(1)$.} $\max_{(\tau,\nu)\in\mathcal{G}}\mathcal{M}_{\tau,\nu}(1)>\gamma$ under $H_0$, whereby we obtain the same $P_{\text{fa}}$-versus-$\gamma$ curve. Since the asymptotic behavior of the class-1 distribution is observed, values of $P_{\text{fa}}$ below $10^{-3}$ can efficiently be estimated by resorting to an extrapolative technique based on the extreme value theory (EVT) \cite{Guida-Iovino-Longo,Weistein}. 
In Figure~\ref{fig_pfa_2}, we report the average number of false detections declared under $H_0$ (shortly, $\text{FD}_0$) versus $P_{\text{fa}}$ when $\Delta_g=T$ (the curves for $\Delta_g=T/2,T/4,T/8$ substantially overlap with the ones for $\Delta_g=T$ and have been omitted for the sake of readability). Clearly, $\text{FD}_0=P_{\text{fa}}$ for the STD. On the other hand, $\text{FD}_0 \geq P_{\text{fa}}$ for the IIC-AMD and MF-PD, as multiple false detections may be declared in the event of a false alarm; in this latter case, however, it is interesting to notice that $\text{FD}_0$ rapidly converges to $P_{\text{fa}}$ as the latter gets smaller, with $\text{FD}_0$ and $P_{\text{fa}}$ becoming practically indistinguishable when $P_{\text{fa}}<0.05$. As a consequence, constraining $\text{FD}_0$ or $P_{\text{fa}}$ is substatially equivalent for all values of practical interest.

\begin{figure*}[!tp]
	\centering
    \ifCLASSOPTIONtwocolumn
	\includegraphics[width=0.9\textwidth]{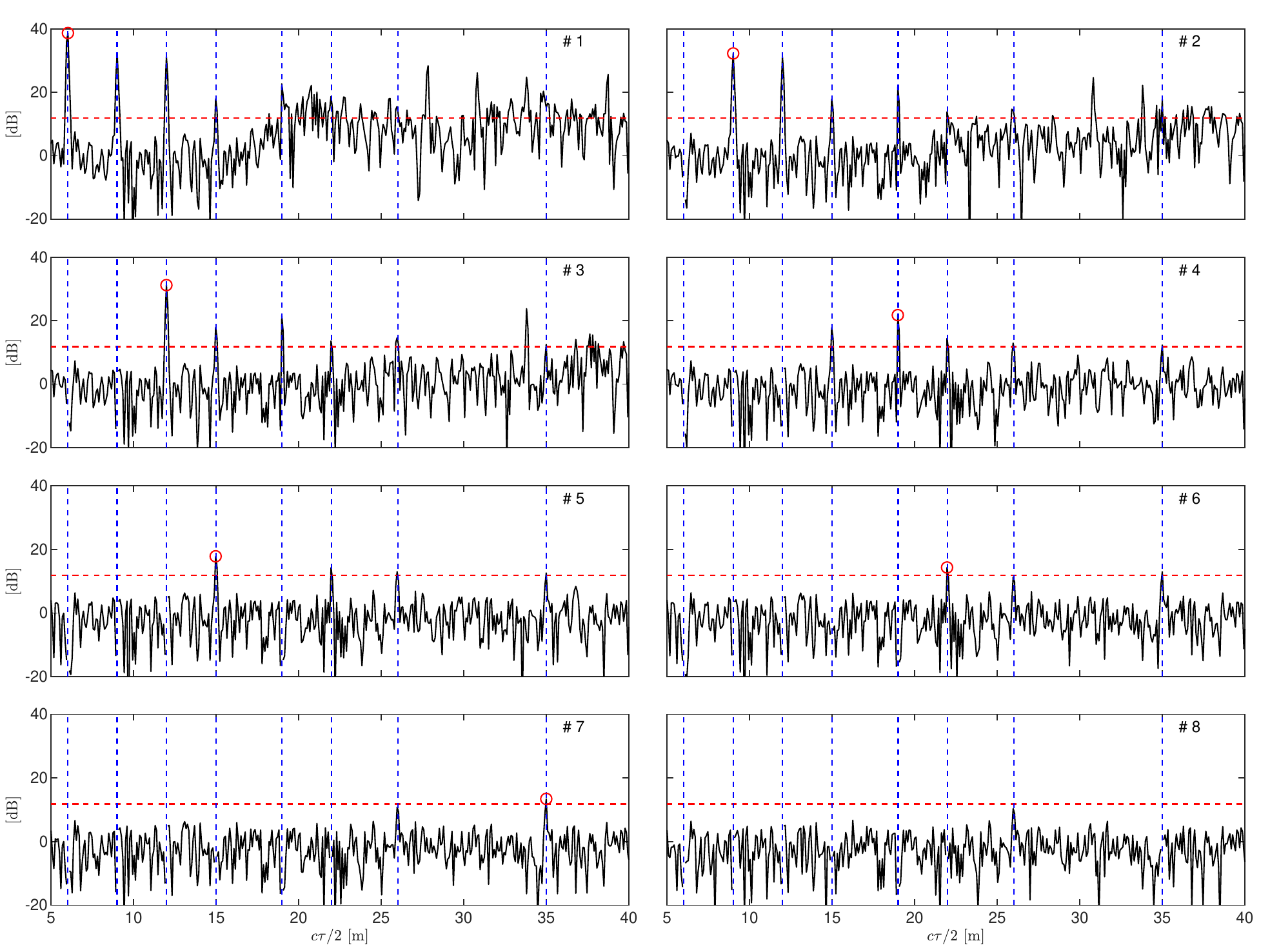}
	\else
	\includegraphics[width=0.9\textwidth]{fig_07_snapshot}
	\fi
	\caption{Evolution of the IIC-AMFD in a single snapshot when $P_{\text{fa}}=10^{-4}$ and $P=8$. The $p$-th subplot (indexed left-to-right and top-to bottom) shows $\mathcal{M}_{\tau,0}(p)$ versus $c\tau/2$ (namely, the inspected range), for $p=1,\ldots,8$. The red dashed horizontal line denotes the detection threshold $\gamma$; the blue dashed vertical lines denote the target positions;  the red circle marker denotes the detection.}
	\label{Fig:ICC-AMFD-snapshot}
\end{figure*}

\begin{figure*}[!tp]
	\centering
	\subfigure[][]{
		\ifCLASSOPTIONtwocolumn
		\includegraphics[width=0.98\columnwidth]{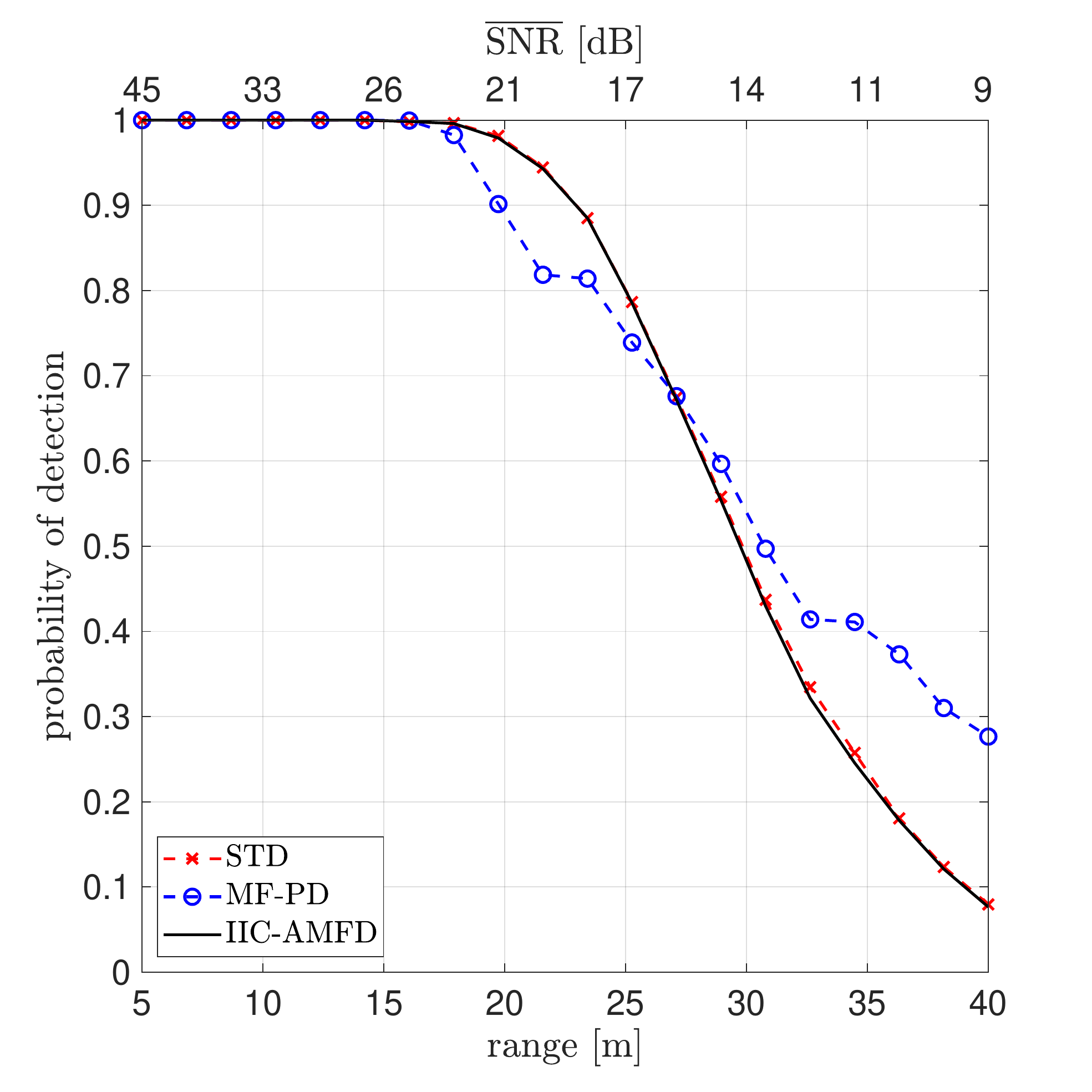}
		\else
		\includegraphics[width=0.46\columnwidth]{fig_11}
		\fi
		\label{fig_pd_1}}
	\subfigure[][]{
		\ifCLASSOPTIONtwocolumn
		\includegraphics[width=0.98\columnwidth]{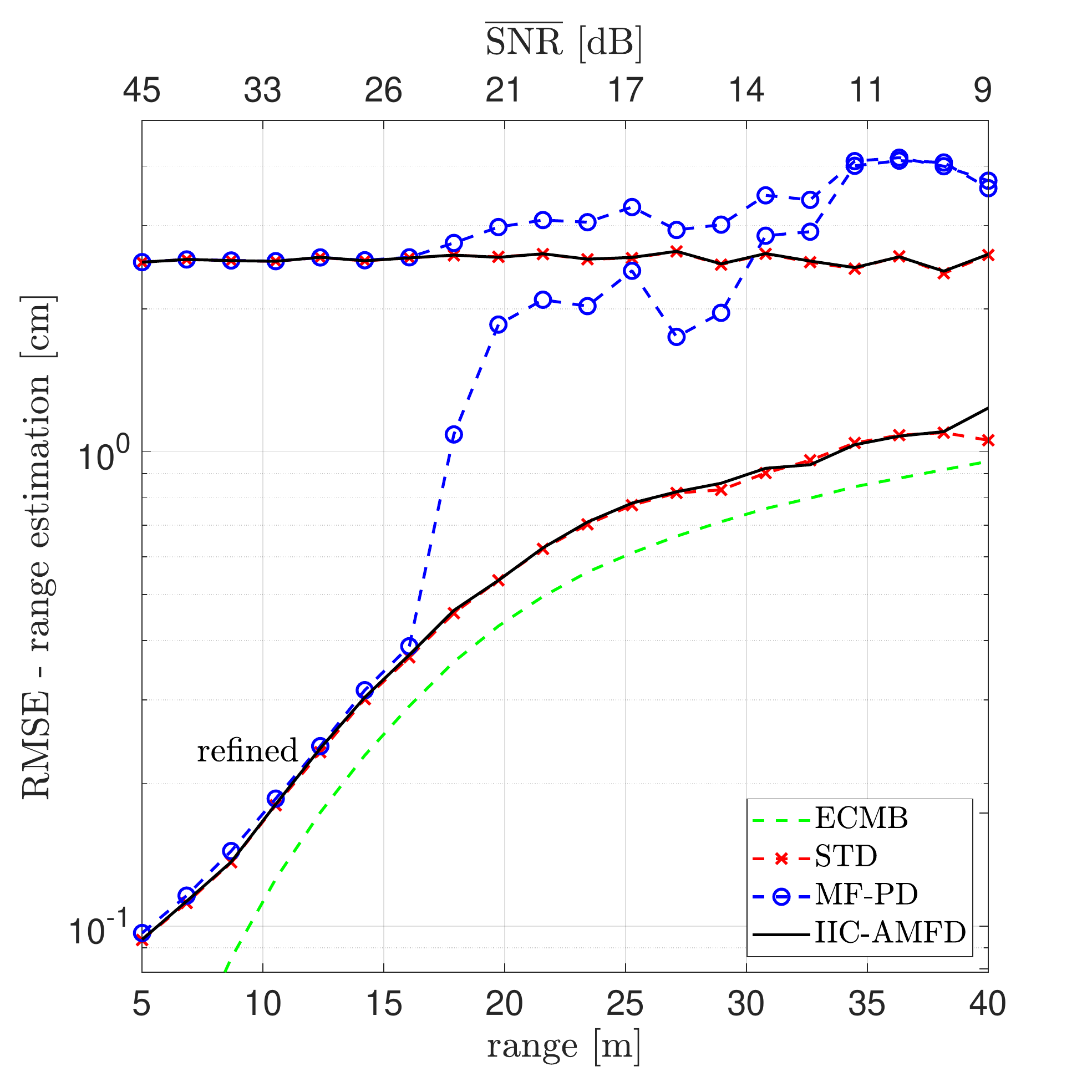}
		\else
		\includegraphics[width=0.46\columnwidth]{fig_12}
		\fi
		\label{fig_pd_2}}
	\subfigure[][]{
		\ifCLASSOPTIONtwocolumn
		\includegraphics[width=0.98\columnwidth]{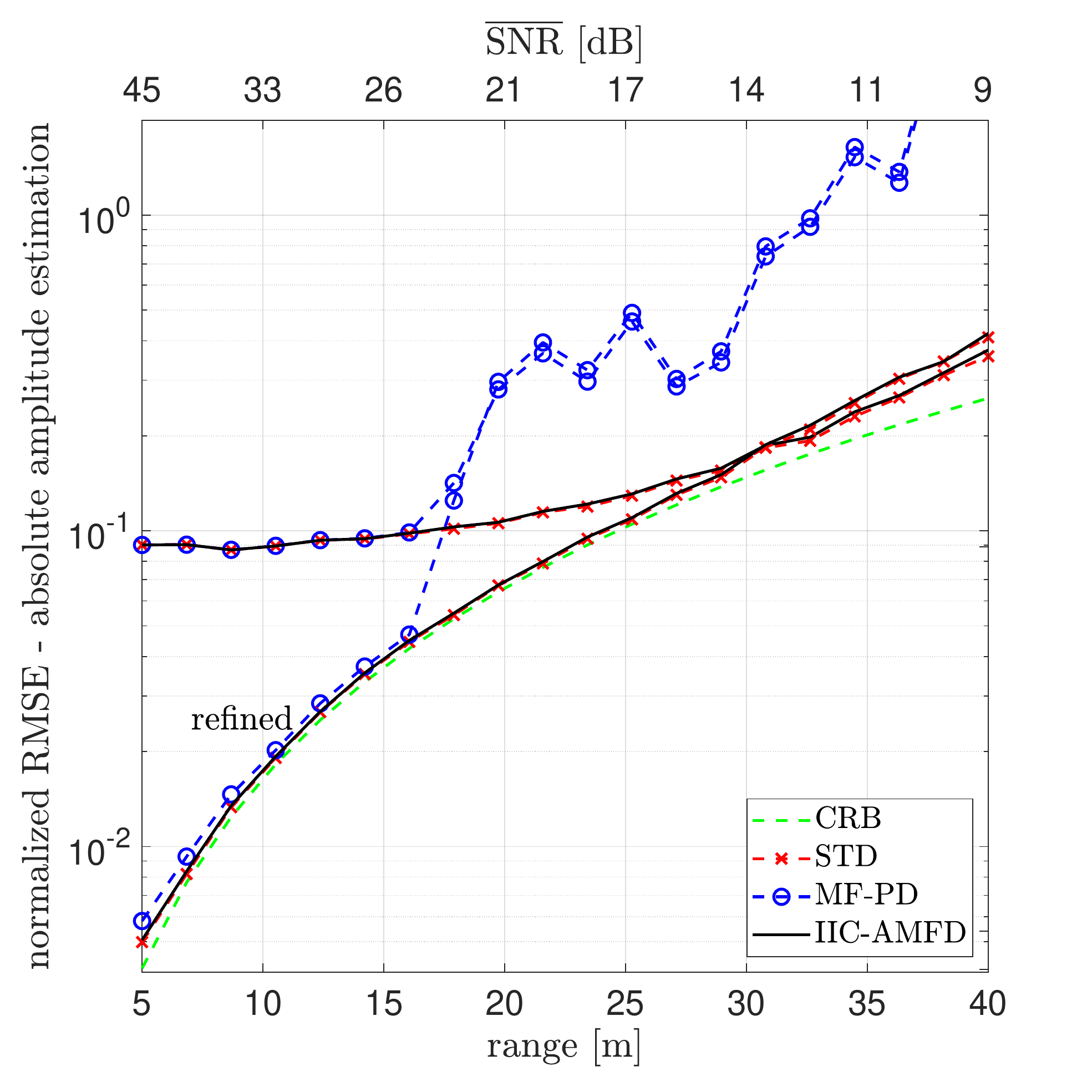}
		\else
		\includegraphics[width=0.46\columnwidth]{fig_13}
		\fi
		\label{fig_pd_3}}
	\subfigure[][]{
		\ifCLASSOPTIONtwocolumn
		\includegraphics[width=0.98\columnwidth]{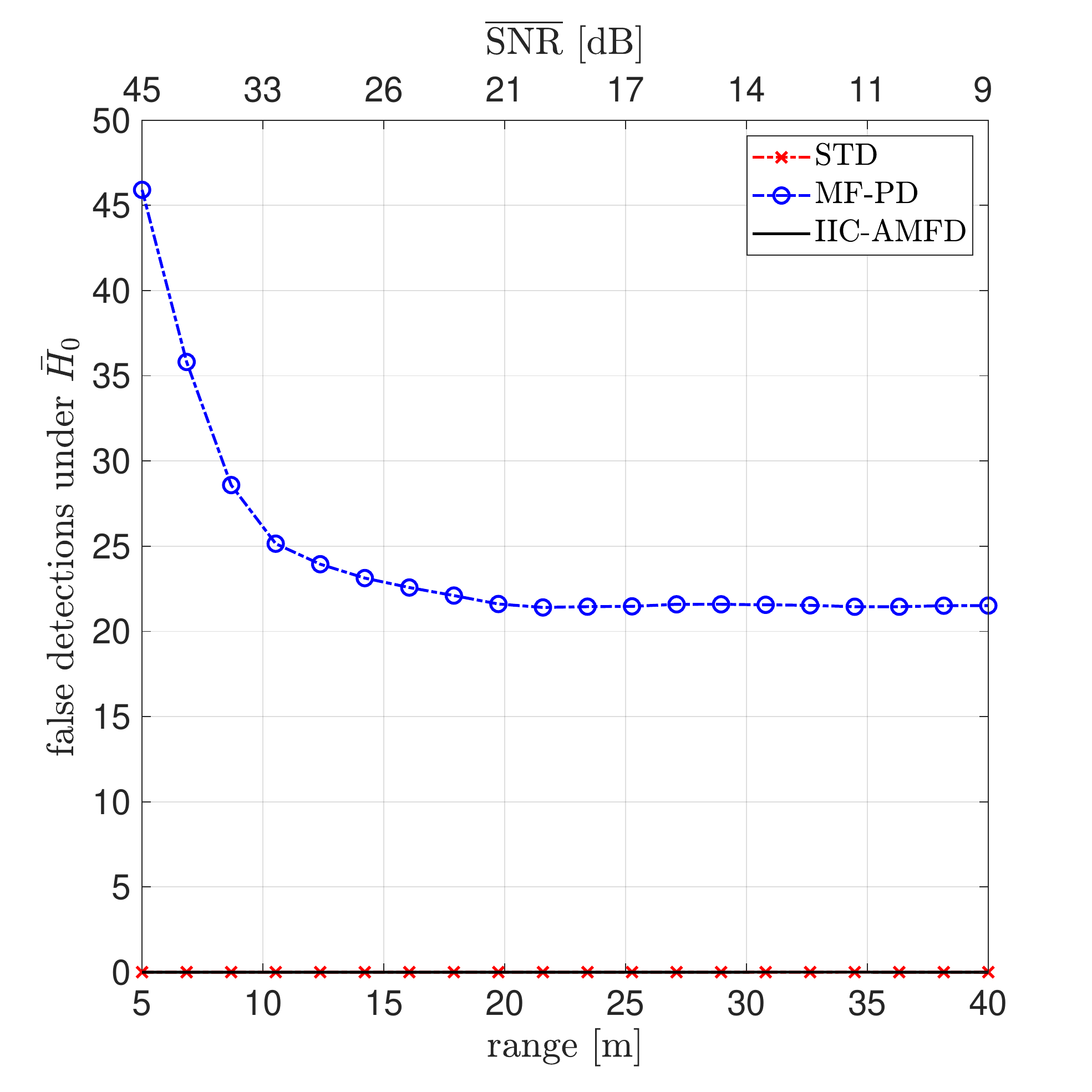}
		\else
		\includegraphics[width=0.46\columnwidth]{fig_14}
		\fi
		\label{fig_pd_4}}
	\caption{Probability of detection (a), \gls{rmse} of the (refined) estimator of the range, normalized \gls{rmse} of the (refined) estimator of the  absolute amplitude (c), and average number of false detections under $\bar{H}_0$ (d)  as a function of the target range when $\Delta_g=T$, $P_{\text{fa}}=10^{-4}$, and $P=8$. }
	\label{fig_pd}
\end{figure*}

In Figure~\ref{Fig:ICC-AMFD-snapshot}, we show the evolution of the proposed IIC-AMFD in a single snapshot when $P_{\text{fa}}=10^{-4}$ and $P=8$ targets are present in the scene; the \gls{rcs} of each target takes a random value in the interval $[0.05,0.2]$ $\text{m}^2$. For the $p$-th iteration, for $p=1,\ldots,8$, we plot the scoring metric  $\mathcal{M}_{\tau,0}(p)$ versus the inspected range (namely, $c\tau/2$). The red dashed horizontal line denotes the detection threshold $\gamma$, while the blue dashed vertical lines denote the target positions. At the first iteration of the algorithm, the scoring metric $\mathcal{M}_{\tau,0}(1)$ coincides with that of the GLRT in~\eqref{eq:GLRT-single-target}. Multiple spurious peaks occurs here as a consequence of the signal spillover at ranges different from those occupied by a target; hence, a  peak detector operating on $\mathcal{M}_{\tau,0}(1)$ would generate many false detections---more on this in Figure~\ref{fig_pd_2}---; in order to overcome this drawback, the IIC-AMFD just detects at this stage the highest peak (indicated by the red circle marker), which is likely to correspond to the target with the largest absolute amplitude (in this case, the closest one). At the second iteration, after removing the interference generated from the first detected target, the highest peak of $\mathcal{M}_{\tau,0}(2)$ is detected, which is likely to correspond to the target with the second largest absolute amplitude (in this case, the second closest one). The effect of removing the interference caused by the first detected target is visible here in the behavior of $\mathcal{M}_{\tau,0}(2)$, which now presents a notch at the range occupied by the first detected target and, more importantly, a much smaller number of spurious secondary peaks. As we proceed with the next iterations, all targets except the weakest one located at $26$ m are detected; also, no false detection is present. At iteration 8, the procedure is stopped as there is no threshold crossing.

\begin{figure*}[!tp]
	\subfigtopskip=-2cm
	\centering
	\subfigure[][]{
		\ifCLASSOPTIONtwocolumn
		\includegraphics[width=0.98\columnwidth]{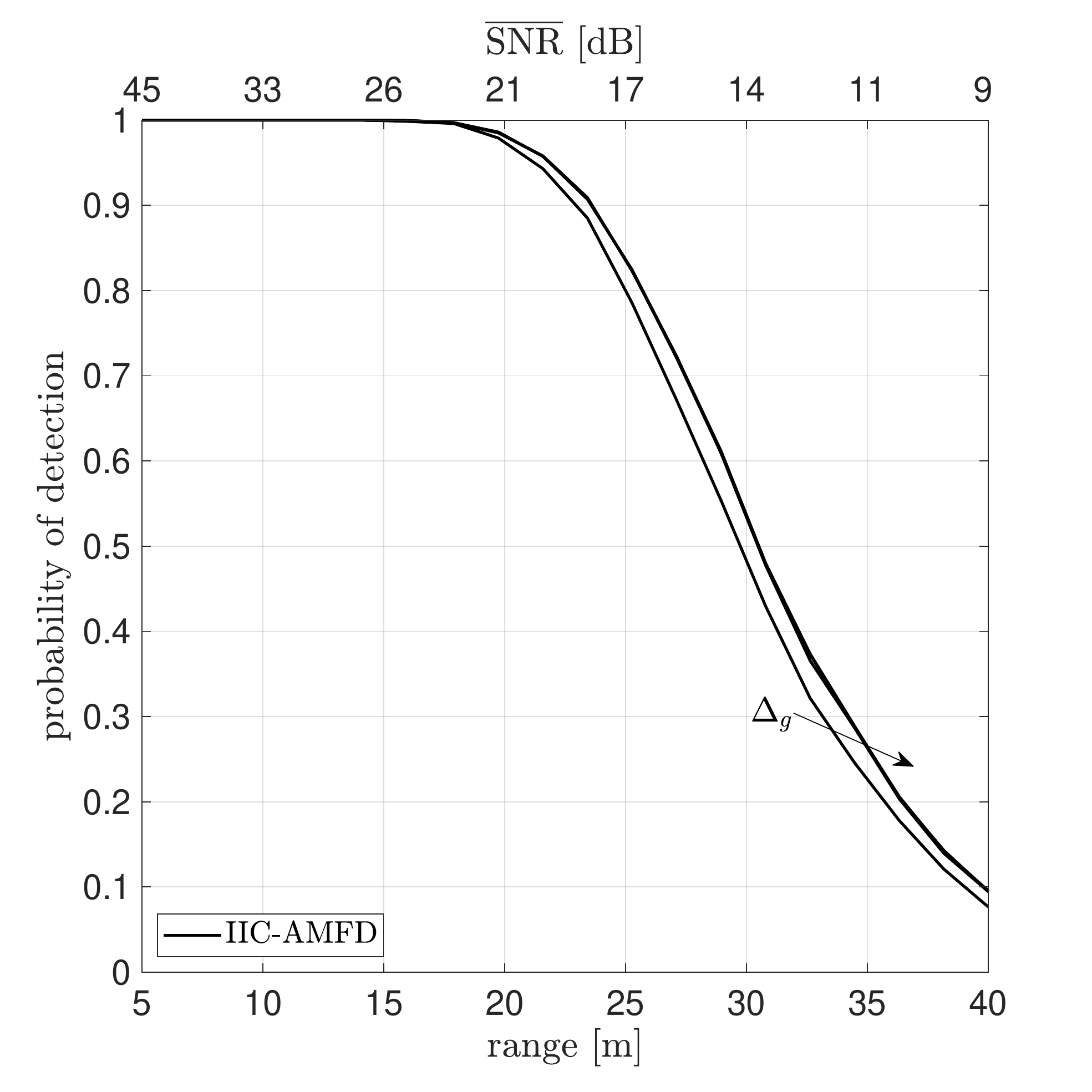}
		\else
		\includegraphics[width=0.46\columnwidth]{fig_15}
		\fi
		\label{fig_pd_ref_1}}
	\subfigure[][]{
		\ifCLASSOPTIONtwocolumn
		\includegraphics[width=0.98\columnwidth]{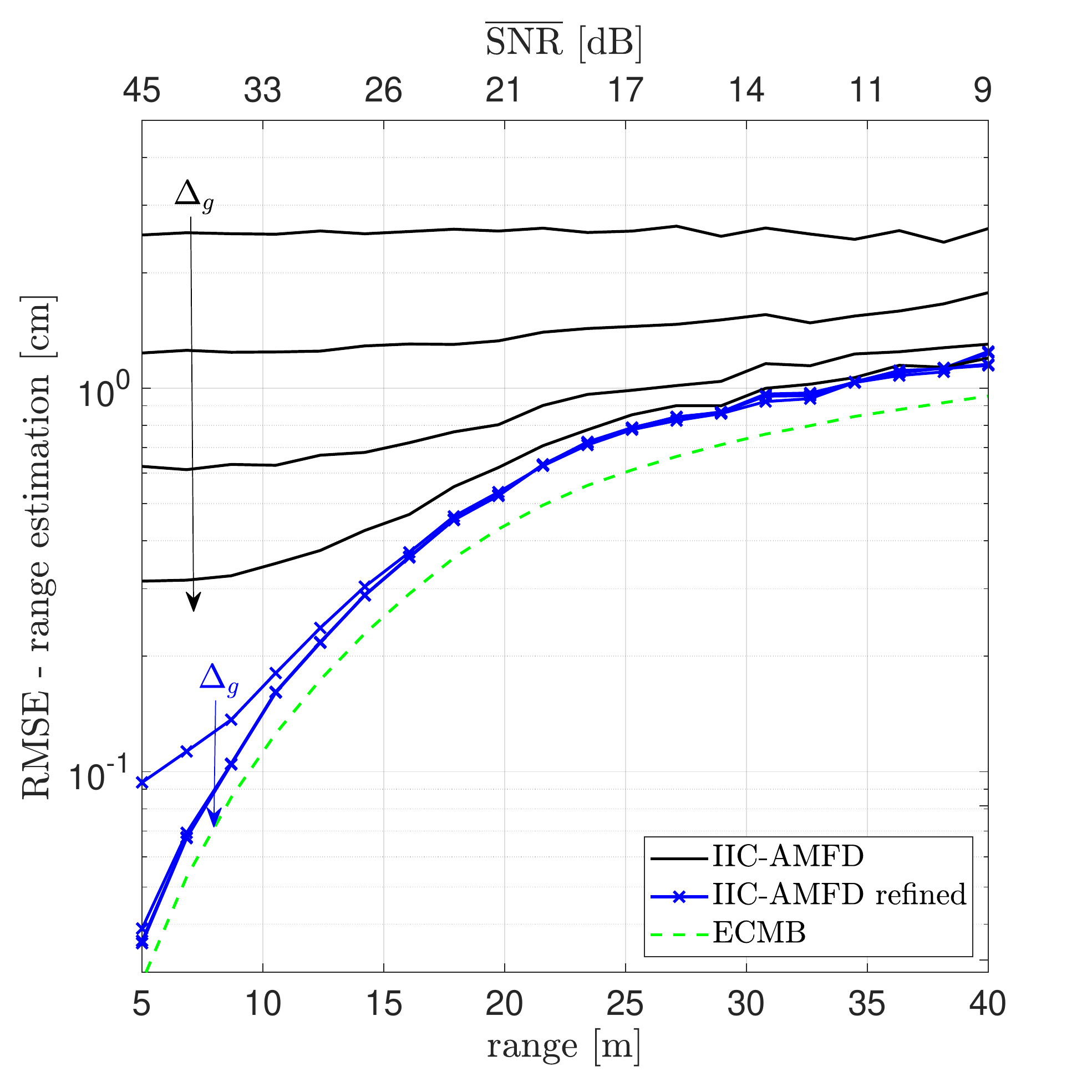}
		\else
		\includegraphics[width=0.46\columnwidth]{fig_16}
		\fi
		\label{fig_pd_ref_2}}
	\caption{Probability of detection (a) and \gls{rmse} of  the (refined) estimator of the range (b) as a function of the target range when $P_{\text{fa}}=10^{-4}$, $P=8$, and $\Delta_g=T,T/2,T/4,T/8$. $\Delta_g$ decreases in the direction of the arrow. }
	\label{fig_pd_ref}
\end{figure*}

In Figure~\ref{fig_pd}, we study the detection and estimation performance as a function of the range, when $\Delta_g=T$ and $P_{\text{fa}}=10^{-4}$. To this end, we consider a reference target  with a \gls{rcs} of $\zeta=0.1$ $\text{m}^2$, whose range is varied from $5$ to $40$ m. In order to assess the performance of the STD, only the reference target has been generated. For the analysis of the MF-PD and ICC-AMFD, instead, seven additional targets (whereby $P=8$) are randomly displaced in the inspected area with a minimum mutual separation of $40$ cm and a \gls{rcs} randomly chosen from the set $[0.05,0.2]$ $\text{m}^2$. We simulate $2000$ independent snapshots. 
Under this setup, Figure~\ref{fig_pd_1} shows the probability of detection of the reference target, defined as the probability that there is a detection with an estimated range differing of at most $cT/2$ (approximatively, $8.5$ cm) from the true range; for the reader sake, we report both the range of the reference target (bottom  $x$-axis) and the average received \gls{snr} (top $x$-axis), the latter being defined as
\begin{equation}
\overline{\text{SNR}}=512 \frac{\mathcal{P}\rm{E}[|\alpha|^2]}{2 W F_{u}\sigma^{2}_{u}}
\end{equation}
where $\alpha$ is the amplitude of the reference target and the factor $512$ is the coherent processing gain granted by the adopted processing window.  Figure~\ref{fig_pd_2} shows the \gls{rmse} of the (refined) estimator of the range, defined as
\begin{equation}
\left(\text{E}\left[|\hat{r}-r|^2\;\rvert\;\text{a detection has occurred}\right]\right)^{1/2}
\end{equation}
where $r$ and $\hat{r}$ are the true range and its (refined) estimate. As a benchmark, we compute the \gls{crb} derived in~\cite{Grossi-TSP-2018}; since this bound depends on the value of $|\alpha|$ observed in each snapshot, we plot its average over all snapshots where a detection is declared, in keeping with the extended Miller-Chang bound (EMCB)~\cite{Miller-Chang-1978,Gini-2000,Leus-2009}. Figure~\ref{fig_pd_3} shows the normalized \gls{rmse} of the (refined) estimator of the absolute amplitude, defined as
\begin{equation}
\frac{\left(\text{E}\left[\left(|\hat{\alpha}|-|\alpha|\right)^2\;\rvert\;\text{a detection has occurred}\right]\right)^{1/2}}
{
\text{E}\left[|\alpha|\right]
}
\end{equation}
where $\alpha$ and $\hat{\alpha}$ are the true amplitude and its (refined) estimate; as a benchmark, we also include the \gls{crb} derived in~\cite{Grossi-TSP-2018}. Finally, Figure~\ref{fig_pd_4} shows the average number of false detections (i.e., detections that cannot be associated with any of the targets present in the scene) as a function of the range of the reference target. 

Several remarks are now in order. First notice that the ICC-AMFD provides detection and estimation performances very close to those of the STD, confirming its robustness with respect to the presence of other targets in the scene. In particular, we obtain a probability of detection larger than 0.8 up to $25$ m; also, using a search grid with $\Delta_g=T$ is already sufficient to provide a range accuracy of about $2.5$ cm; the estimation error is farther reduced by using the refined estimator proposed in Section~\ref{Sec:refined-estimate} and an RMSE of few millimeters---quite close to the EMCB---can be obtained at very short distances. In the generated snapshots, no false detection has been produced by both the ICC-AMFD and the STD. On the other hand, the MF-PD has produced many false detections in each snapshot (see also the upper-left subplot of Figure~\ref{Fig:ICC-AMFD-snapshot}). These false detections are not caused by spikes of the underlying noise (as we are operating at $P_{\text{fa}}=10^{-4}$), but from the signal leakage at look directions where no target is actually present; in particular, the number of false detections increases when the reference target is closer, as a stronger echo produces a more severe spillover.

Figures~\ref{fig_pd_ref_1} and~\ref{fig_pd_ref_2} compare the probability of detection and the RMSE of the (refined) estimator of the range, respectively, for $\Delta_g=T,T/2,T/4,T/8$. The same setup considered in Figure~\ref{fig_pd} is adopted here. The choice of $\Delta_g$ has marginal impact on the probability of detection, which only present a minor increment form $\Delta_g=T$ to $\Delta_g=T/2$ and, then, saturates. On the other hand, decreasing $\Delta_g$ helps improving the accuracy of the initial range estimate; however, it has no significant effect on the refine range estimate.

\begin{figure*}[!tp]
	\subfigtopskip=-2cm
	\centering
	\subfigure[][]{
		\ifCLASSOPTIONtwocolumn
		\includegraphics[width=0.98\columnwidth]{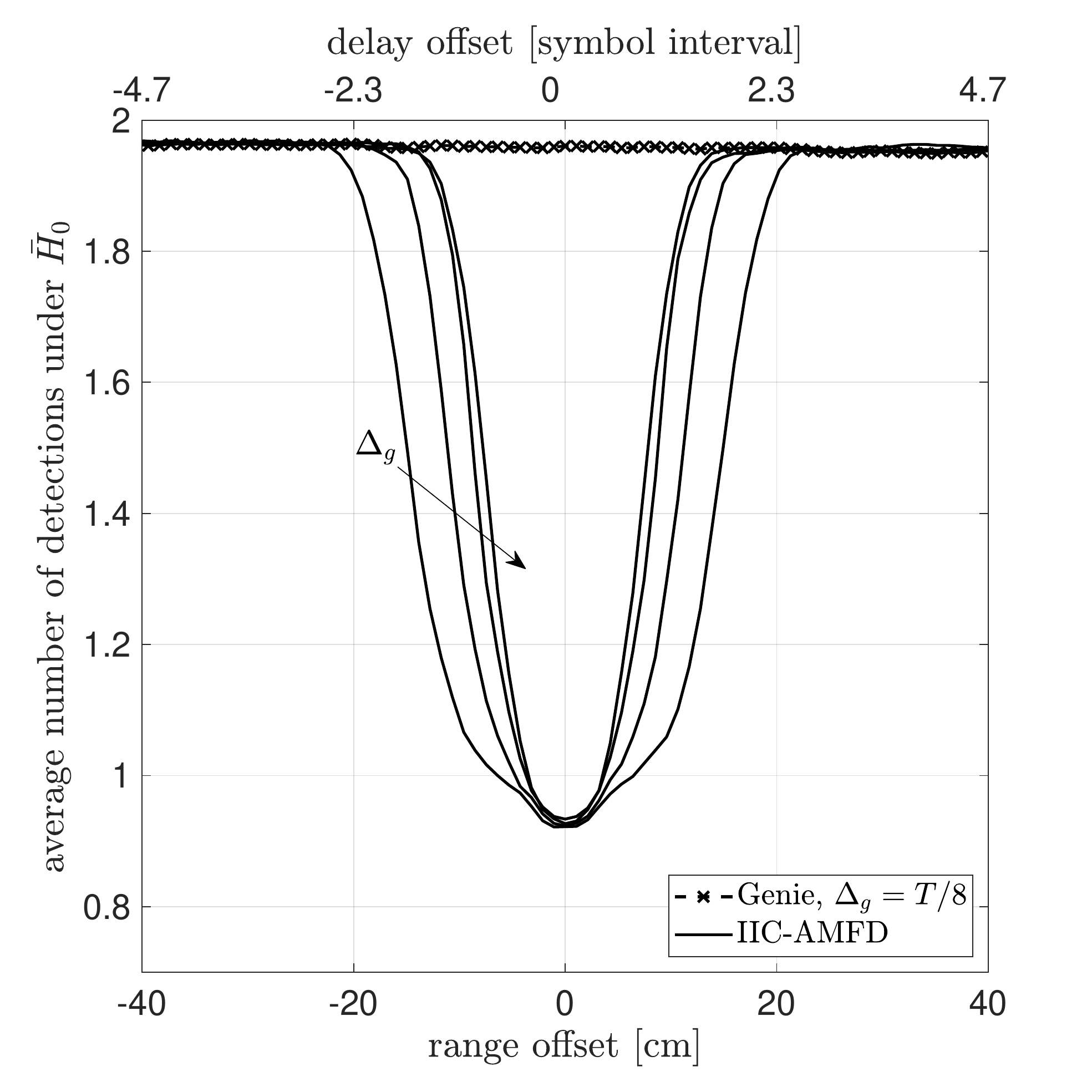}
		\else
	    \includegraphics[width=0.46\columnwidth]{fig_17}
		\fi
		\label{fig_pd_close_1}}
	\subfigure[][]{
		\ifCLASSOPTIONtwocolumn
			\includegraphics[width=0.98\columnwidth]{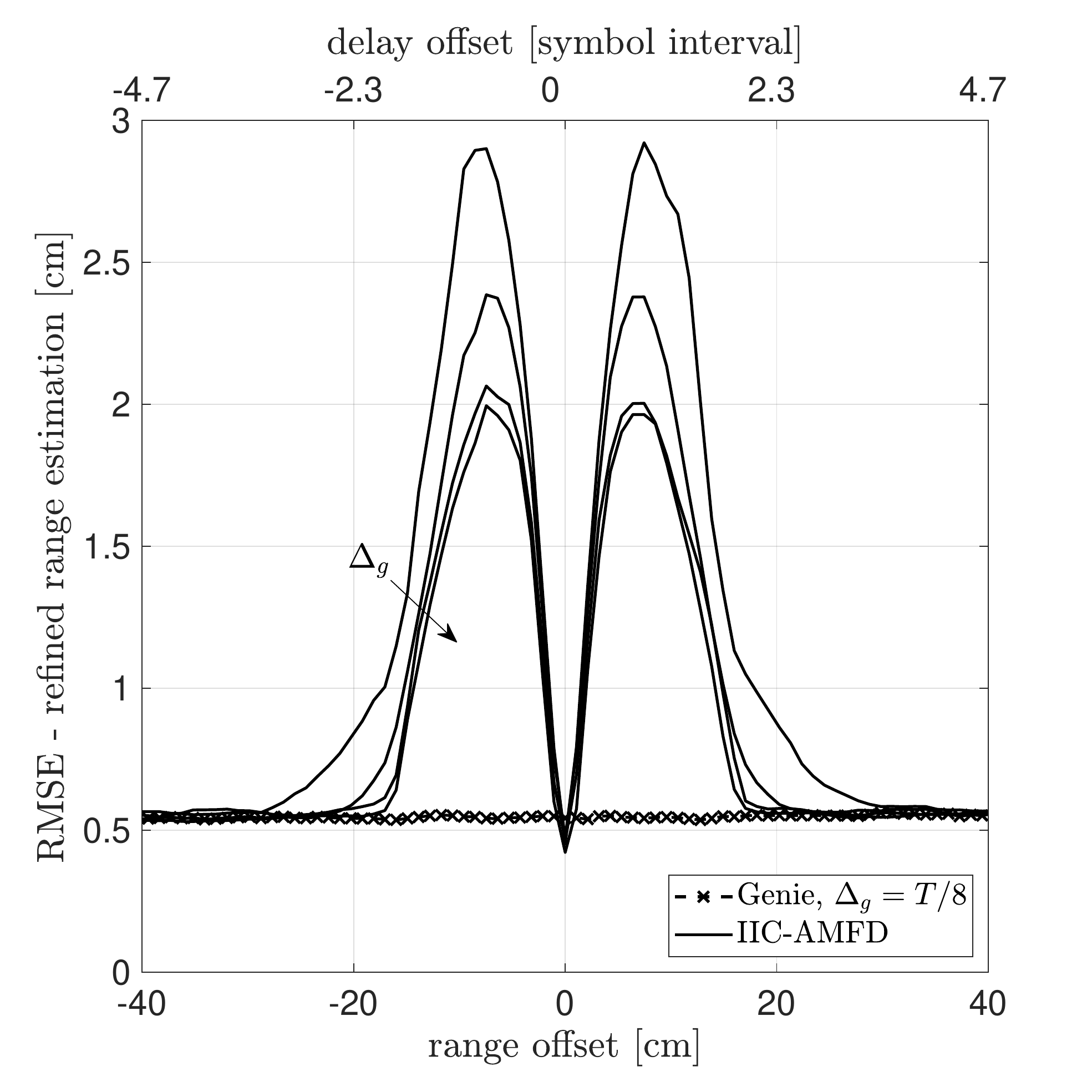}
		\else
			\includegraphics[width=0.46\columnwidth]{fig_18}
		\fi
		\label{fig_pd_close_2}}
	\caption{Average number of detections (a) and \gls{rmse} of the refined estimator of the range (b) as a function of the mutual target distance when $P_{\text{fa}}=10^{-4}$, $P=2$, and $\Delta_g=T,T/2,T/4,T/8$. $\Delta_g$ decreases in the direction of the arrow.}
	\label{fig_pd_close}
\end{figure*}

To farther investigate the impact of $\Delta_g$ on the performance of the ICC-AMFD, we consider now a different experiment. We simulate two targets in the scene with a \gls{rcs} of 0.1 $\text{m}^2$: one at a range of $20$ m and another in its close proximity. Figures~\ref{fig_pd_close_1} and~\ref{fig_pd_close_2} report the average number of detections in each snapshot and the RMSE of the refined estimator of the range, respectively, as a function of the mutual target distance (bottom $x$-axis) for $\Delta_g=T,T/2,T/4,T/8$. On the top $x$-axis we also report  the corresponding delay offset normalized by the symbol interval. For the sake of comparison, we include the performance of a genie detector which operates as follows. At first, it perfectly removes the echo from the first target and uses the STD to detect the other echo; then, it perfectly removes the echo from the second target and uses the STD to detect the other echo. It is seen that reducing $\Delta_g$ is helpful to better resolve close targets. In particular, moving from  $\Delta_g=T$ to $\Delta_g=T/8$ improves the range resolution of about $8.5$ cm (corresponding to a delay of about one symbol interval); notice that, when the delay separation drops below $2T=1/W$, then the average number of detections rapidly decreases (see also the previous discussion in Section~\ref{SEC:Range-Doppler-accuracy}), as one of the targets may be masked by the other one and, therefore, missed. Finally, notice that the range accuracy first degrades when the targets get closer, as a consequence of the mutual interference; then, when the separation gets smaller than $8.5$ cm, it improves as the two targets are essentially seen as a unique object.

\section{Conclusions}\label{SEC:conclusions}
In this work, we have considered an opportunistic radar which exploits the \gls{sls} phase of IEEE 802.11ad communication standard operating at mmWaves and we have studied the problem of multiple target detection and localization. We have shown that the imperfect auto-correlation of the probing signal prevents the use a simple matched-filter receiver, as the signal leakage may generate false detections and/or target masking. Also, we have proposed a novel adaptive procedure, which extracts and detects the prospective echoes one-by-one from the received signal: the main idea here is to adaptively remove the interference caused by the previously detected targets. The numerical analysis has shown that the proposed solution grants detection and estimation performances very close to those obtained in a single-target scenario for the same probability of false alarm. This result comes at the price of an implementation complexity cubic (rather than linear, as for the matched filter) with the number of symbol intervals spanned by the processing window and linear with the number of targets. However, a wise choice of the processing window allows to achieve a satisfactory coverage at an affordable cost: specifically, processing a data segment spanning only 512 symbol intervals is already sufficient to have a probability of detection larger than 0.8 up to $25$ m and a range accuracy of few centimeters for reasonable system parameters and targets with a \gls{rcs} of 0.1 m$^2$. Also, the proposed refined estimator can farther reduce the range accuracy to few millimeters at shorter distances. Accurate estimation of the target amplitude is also possible. On the other hand, the short duration of the probing signal prevents the measurement of the range-rate. We underline that processing longer segments of the \gls{cphy} packet, albeit in principle advantageous from the point of view of the achievable performance, may turn out to be computationally prohibitive for practical applications.

Future works along this research line should possibly investigate the problem of detecting range-spread objects and jointly estimating their position and extension; also, they should consider the effect of a more accurate channel model, wherein multi-path propagation may possibly result into ghost targets.

\section{Acknowledgment}
The work of E. Grossi and L. Venturino was supported
by the MIUR program “Dipartimenti di Eccellenza 2018--2022".

\bibliographystyle{IEEEtran} 
\bibliography{manuscript}

\end{document}

%% file: acronyms.tex
\newacronym{rcs}{RCS}{radar cross-section}
\newacronym{amfd}{AMFD}{adaptive matched filter detector}
\newacronym{mf}{MF}{matched filter}
\newacronym{mf-pd}{MF-PD}{matched-filter peak detector}
\newacronym{wmf}{WMF}{whitening matched filter}
\newacronym{br}{BR}{beam refinement}
\newacronym{iss}{ISS}{initiator sector sweep}
\newacronym{rss}{RSS}{responder sector sweep}
\newacronym{ssw-fb}{SSF}{sector sweep feedback}
\newacronym{ssw-ack}{SSA}{sector sweep acknowledgement}
\newacronym{iaa}{IAA}{iterative adaptive approach}
\newacronym{iaa-apes}{IAA-APES}{iterative adaptive approach for amplitude and phase estimation}
\newacronym{iic}{IIC}{iterative interference cancellation}
\newacronym{bpsk}{BPSK}{binary phase shift keying}

\newacronym{mac}{MAC}{multiple-access channel}
\newacronym{bc}{BC}{broadcast channel}
\newacronym{mimo}{MIMO}{multiple-input multiple-output}
\newacronym{siso}{SISO}{single-input single-output}
\newacronym{af}{AF}{amplify-and-forward}
\newacronym{df}{DF}{decode-and-forward}
\newacronym{cf}{CF}{compress-and-forward}
\newacronym{mwrc}{MWRC}{multi-way relay channel}
\newacronym{dmmwrc}{DM-MWRC}{discrete memoryless multi-way relay channel}
\newacronym{pde}{PDE}{partial data exchange}
\newacronym{fde}{FDE}{full data exchange}
\newacronym{iid}{i.i.d.\@}{independent and identically distributed}
\newacronym{awgn}{AWGN}{additive white Gaussian noise}
\newacronym{awg}{AWG}{additive white Gaussian}
\newacronym{sic}{SIC}{successive interference cancellation}
\newacronym{snr}{SNR}{signal-to-noise ratio}
\newacronym{sinr}{SINR}{signal-to-interference-plus-noise ratio}
\newacronym{inr}{INR}{interference to noise ratio}
\newacronym{zf}{ZF}{zero-forcing}
\newacronym{mmse}{MMSE}{minimum mean square error}
\newacronym{sud}{SUD}{single user decoding}
\newacronym{dof}{DoF}{degrees of freedom}
\newacronym{gdof}{GDoF}{generalized degrees of freedom}
\newacronym{nnc}{NNC}{noisy network coding}
\newacronym{dmn}{DMN}{discrete memoryless network}
\newacronym{csi}{CSI}{channel state information}
\newacronym{pmf}{pmf}{probability mass function}
\newacronym{dmic}{DM-IC}{discrete memoryless interference channel}
\newacronym{ic}{IC}{interference channel}
\newacronym{ee}{EE}{energy efficiency}
\newacronym{gee}{GEE}{global energy efficiency}
\newacronym{ian}{IAN}{treating interference as noise}
\newacronym{snd}{SND}{simultaneous non-unique decoding}
\newacronym{hk}{HK}{Han-Kobayashi}
\newacronym{rf}{RF}{radio frequency}
\newacronym{pa}{PA}{power amplifier}
\newacronym{lna}{LNA}{low noise amplifier}
\newacronym{lo}{LO}{local oscillator}
\newacronym{adc}{ADC}{analog-to-digital converter}
\newacronym{dac}{DAC}{digital-to-analog converter}
\newacronym{dsp}{DSP}{digital signal processing}
\newacronym{brd}{BRD}{best response dynamics}
\newacronym{ne}{NE}{Nash equilibrium}
\newacronym{lhs}{LHS}{left-hand side}
\newacronym{rhs}{RHS}{right-hand side}
\newacronym{glrt}{GLRT}{generalized likelihood ratio test}
\newacronym{lr}{LR}{likelihood ratio}
\newacronym{llr}{LLR}{log-LR}
\newacronym{ml}{ML}{maximum likelihood}
\newacronym{los}{LOS}{line of sight}
\newacronym{sls}{SLS}{sector level sweep}
\newacronym{cphy}{CPHY}{control physical}
\newacronym{qo}{QO}{quasi omni-directional}
\newacronym{ssw}{SSW}{sector sweep}
\newacronym{rmse}{RMSE}{root mean square error}
\newacronym{crb}{CRB}{Cram\'er-Rao bound}
\newacronym{cdf}{CDF}{cumulative distribution function}
\newacronym{pdf}{PDF}{probability density function}